\documentclass[%
    aps,
    pra,
    reprint,
    10pt,
    superscriptaddress,
    amsfonts,
    amssymb,
    amsmath,
    preprintnumbers,
    noshowpacs,
    tightenlines,
    floats,
    notitlepage,
    final,
    letterpaper,
    oneside,
    twocolumn,
    nofootinbib,
    longbibliography,
    ]{revtex4-2}
\pdfoutput=1
\usepackage[utf8]{inputenc}
\usepackage[T1]{fontenc}
\usepackage[american]{babel}
\usepackage[british,cleanlook,printdayon]{isodate}
\usepackage[Symbolsmallscale]{upgreek}
\usepackage{mathtools}
\usepackage{isomath}
\usepackage[protrusion,tracking,kerning,spacing,final,babel]{microtype}
\usepackage{physics}
\usepackage[final]{hyperref}
\hypersetup{%
    colorlinks=true,
    citecolor=blue,
    linkcolor=blue,
    urlcolor=blue,
    }%
\usepackage{cleveref}
\usepackage{orcidlink}

\newcommand{\rme}{\mathrm{e}}
\newcommand{\rmi}{\mathrm{i}}
\newcommand{\down}{\downarrow}
\newcommand{\up}{\uparrow}

\begin{document}

\title{Bubbles in a box:\\Eliminating edge nucleation in cold-atom simulators of vacuum decay}

\author{Alexander~C.~Jenkins\,\orcidlink{0000-0003-1785-5841}}
\altaffiliation{Corresponding author}
\email{acj46@cam.ac.uk}
\affiliation{Kavli Institute for Cosmology, University of Cambridge, Madingley Road, Cambridge CB3 0HA, UK}
\affiliation{DAMTP, University of Cambridge, Wilberforce Road, Cambridge CB3 0WA, UK}

\author{Hiranya~V.~Peiris\,\orcidlink{0000-0002-2519-584X}}
\affiliation{Kavli Institute for Cosmology, University of Cambridge, Madingley Road, Cambridge CB3 0HA, UK}
\affiliation{Institute of Astronomy, University of Cambridge, Madingley Road, Cambridge CB3 0HA, UK}
\affiliation{The Oskar Klein Centre for Cosmoparticle Physics, Department of Physics, Stockholm University, AlbaNova, Stockholm SE-106 91, Sweden}

\author{Andrew~Pontzen\,\orcidlink{0000-0001-9546-3849}}
\affiliation{Institute for Computational Cosmology, Department of Physics, Durham University, South Road, Durham, DH1 3LE, UK}

\begin{abstract}
    The decay of metastable `false vacuum' states via bubble nucleation plays a crucial role in many cosmological scenarios.
    Cold-atom analog experiments will soon provide the first empirical probes of this process, with potentially far-reaching implications for early-Universe cosmology and high-energy physics.
    However, an inevitable difference between these analog systems and the early Universe is that the former have a boundary.
    We show, using a combination of Euclidean calculations and real-time lattice simulations, that these boundaries generically cause rapid bubble nucleation on the edge of the experiment, obscuring the bulk nucleation that is relevant for cosmology.
    We demonstrate that implementing a high-density `trench' region at the boundary completely eliminates this problem, and recovers the desired cosmological behavior.
    Our findings are relevant for ongoing efforts to probe vacuum decay in the laboratory, providing a practical solution to a key experimental obstacle.
\end{abstract}

\selectlanguage{british}
\date{\today}
\maketitle
\selectlanguage{american}

%%%%%%%%%%%%%%%%%%%%%%%%%%%%%%%%%%%%%%%%%%%%%%%%%%%%%%%%%%%%%%%%%%%%%%%%%%%%%%%
\section{Introduction}\label{sec:intro}
One of the fundamental challenges of cosmology is that it is an observational science, not an experimental one: one cannot control the system in question (the Universe), and can only access a single realization of it, drawn from an inherently stochastic quantum process.
Reconstructing the underlying physical laws from within this one realization, without any freedom to vary parameters or conduct controlled experiments, is a daunting task.
This problem is particularly acute for the very early Universe, for which observational data are scarce and the underlying physics is poorly understood.
These challenges have driven a surge of interest in simulating early-Universe theories using quantum analog experiments~\cite{Opanchuk:2013lgn,Fialko:2014xba,Fialko:2016ggg,Braden:2017add,Billam:2018pvp,Braden:2019vsw,Billam:2020xna,Ng:2020pxk,Billam:2021qwt,Billam:2021nbc,Billam:2022ykl,Jenkins:2023eez,Jenkins:2023npg,Zenesini:2023afv,Cominotti:2025qia,Fischer:2004bf,Visser:2004qp,Visser:2005ss,Weinfurtner:2006wt,Neuenhahn:2012dz,Su:2014osc,Zache:2017dnz,Eckel:2017uqx,Abel:2020qzm,Chatrchyan:2020cxs,Milsted:2020jmf,Banik:2021xjn,Lagnese:2021grb,Viermann:2022wgw,Tolosa-Simeon:2022umw,Tajik:2022lyt,Darbha:2024srr,Schmidt:2024zpg,Zhu:2024dvz,Schutzhold:2025qna}.
By emulating the behavior of relativistic fields, these analogs enable controllable and reproducible cosmological experiments, with transformative potential for fundamental physics.

Vacuum decay is an emblematic use case for such analogs.
This process, in which a relativistic scalar field decays from a metastable `false vacuum' state by nucleating bubbles of true vacuum~\cite{Coleman:1977py,Callan:1977pt}, is nonperturbative and inherently quantum, such that any analytical description or numerical simulation must resort to assumptions and approximations.
Analog simulations of vacuum decay promise to provide the first empirical tests of these descriptions, potentially revealing interesting new phenomenology (including bubble clustering~\cite{Pirvu:2021roq,DeLuca:2021mlh}, dynamical precursors~\cite{Pirvu:2023plk}, and time-dependent decay rates~\cite{Batini:2023zpi,Pirvu:2024ova,Pirvu:2024nbe}), with implications for inflation~\cite{Guth:2007ng,Aguirre:2007an,Aguirre:2007gy,Feeney:2010jj,Feeney:2010dd}, baryogenesis~\cite{Kuzmin:1985mm,Cohen:1993nk,Morrissey:2012db}, gravitational waves~\cite{Kosowsky:1991ua,Kamionkowski:1993fg,Caprini:2015zlo}, and Higgs metastability~\cite{Ellis:2009tp,Degrassi:2012ry,Buttazzo:2013uya}.

Recent years have seen significant progress toward simulating vacuum decay using ultracold atomic condensates, including theoretical developments in modeling these systems and understanding the regimes in which they behave relativistically~\cite{Opanchuk:2013lgn,Fialko:2014xba,Fialko:2016ggg,Braden:2017add,Billam:2018pvp,Braden:2019vsw,Billam:2020xna,Ng:2020pxk,Billam:2021qwt,Billam:2021nbc,Billam:2022ykl,Jenkins:2023eez,Jenkins:2023npg}, as well as experimental realization of \emph{non}relativistic vacuum decay in inhomogeneous condensates~\cite{Zenesini:2023afv,Cominotti:2025qia}.
The ultimate goal is to build a simulator that ($i$) has a well-defined relativistic regime, and ($ii$) is as homogeneous as possible, to recreate the conditions relevant to early-Universe theories.
Efforts toward this goal are ongoing at the Cavendish Laboratory in Cambridge as part of the QSimFP Consortium,\footnote{\url{https://qsimfp.org/}} using an optical box trap~\cite{Gaunt:2013box,Navon:2021mcf} to ensure homogeneity across the bulk of the condensate.

However, any cold-atom experiment will inevitably be inhomogeneous at its boundary, where the walls of the box force the atomic density to zero.
As we demonstrate below, this is a potentially serious problem for analog vacuum decay, as these inhomogeneities generically catalyze rapid bubble nucleation on the boundary of the experiment, obscuring the bulk nucleation that is of cosmological interest.
This accelerated decay was previously observed numerically in Ref.~\cite{Billam:2022ykl}, and is closely related to the phenomenon of seeded decay from impurities in the bulk~\cite{Moss:1984zf,Gregory:2013hja,Billam:2018pvp,Caneletti:2024kww}.

In this paper, we show that edge nucleation can be eliminated by engineering the trapping potential to create a `trench' of high atomic density at the boundary.
We demonstrate this analytically in Sec.~\ref{sec:edges} using Euclidean calculations in the thin-wall regime, and verify it beyond this regime in Sec.~\ref{sec:sims} using real-time semiclassical lattice simulations.
Our focus is on quantum nucleation in the Rabi-coupled system described in, e.g., Refs.~\cite{Jenkins:2023eez,Jenkins:2023npg}; a companion paper~\cite{Brown:2025wxy} uses alternative numerical techniques to investigate thermal nucleation in three different analog systems.
In all cases studied, engineering the potential allows one to completely eliminate edge nucleation.

%%%%%%%%%%%%%%%%%%%%%%%%%%%%%%%%%%%%%%%%%%%%%%%%%%%%%%%%%%%%%%%%%%%%%%%%%%%%%%%
\section{Edges in the analog false vacuum}\label{sec:edges}
We begin by reviewing the analog system studied in Refs.~\cite{Jenkins:2023eez,Jenkins:2023npg} and the Euclidean description of bulk nucleation in the thin-wall limit~\cite{Coleman:1977py,Callan:1977pt}.
We then consider edge nucleation in this limit, showing that this is exponentially enhanced in a standard box trap, before demonstrating how a high-density boundary layer eliminates this problem.
Our treatment here follows Ref.~\cite{Caneletti:2024kww}, where similar calculations were used to study the seeding of bubbles by impurities in the bulk.
We give only a brief overview of the analog system, focusing on the details that are necessary to understand the edge nucleation problem and its solution.
Further details can be found in Appendix~\ref{sec:tensions}.

%-----------------------------------------------------------------------------%
\subsection{The relativistic analog}
\begin{figure}[t!]
    \centering
    \includegraphics{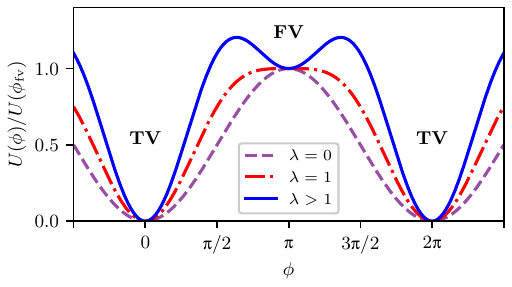}
    \caption{\label{fig:phi-potential}
    Effective self-interaction potential for the relative phase $\phi=\phi_\down-\phi_\up$, cf.~Eq.~\eqref{eq:phi-potential}.
    The constant part of the Rabi coupling generates a cosine potential (purple dashed curve), while the modulation generates a potential barrier with height proportional to the square of the modulation amplitude.
    This amplitude is set by the dimensionless parameter $\lambda$, which is normalized such that the potential is flat at the points $\phi=\uppi\pmod{2\uppi}$ when $\lambda=1$ (red dot-dashed curve).
    For $\lambda>1$ (blue solid curve) these points become metastable local potential minima (labelled `FV' for false vacuum).
    The global minima are at $\phi=0\pmod{2\uppi}$ for all values of $\lambda$ (labelled `TV' for true vacuum).}
\end{figure}

Our system is a gas of two internal states of a bosonic isotope, labelled $\ket{i}=\ket{\down},\ket{\up}$.
At ultracold temperatures, each species forms a condensate described by a many-body wavefunction $\psi_i=\sqrt{n_i}\exp(\rmi\phi_i)$, with density $n$ and phase $\phi$.
As well as nonlinear interactions due to two-body $s$-wave scattering, the condensates interact via a Rabi coupling (a coherent electromagnetic beam with frequency corresponding to the energy splitting between the two states) whose amplitude is rapidly modulated.
On timescales longer than the modulation period, this generates an effective potential for the relative phase $\phi=\phi_\down-\phi_\up$.
Under suitable experimental conditions, the equation of motion for $\phi$ becomes that of a relativistic scalar field,
    \begin{equation}
    \label{eq:eom}
        (c_\phi^{-2}\partial_t^2-\laplacian)\phi+\dv{U}{\phi}=0,
    \end{equation}
    with a periodic potential,
    \begin{equation}
    \label{eq:phi-potential}
        U(\phi)=\epsilon\frac{m_\phi^2c_\phi^2}{\hbar^2}\qty(1-\cos\phi+\frac{\lambda^2}{2}\sin^2\phi),
    \end{equation}
    as illustrated in Fig.~\ref{fig:phi-potential}.
Here $c_\phi$ is the sound speed of the $\phi$-phonons, which corresponds to the speed of light in the effective relativistic theory, while $m_\phi$ is a mass scale, comparable to the atomic mass $m$.
The dimensionless constants $\epsilon,\lambda$ are associated with the mean amplitude and modulation amplitude of the Rabi coupling, respectively.
The effective relativistic equation of motion~\eqref{eq:eom} is valid only in the regime where the density fluctuations $\updelta n_\down$, $\updelta n_\up$ sourced by the relative phase dynamics are perturbatively small, which in turn requires $U(\phi)$ to be much smaller than the overall energy density of the system.
Since the latter is on the order of $\sim m_\phi^2c_\phi^2/\hbar^2$, we see from Eq.~\eqref{eq:phi-potential} that this requirement is satisfied when
    \begin{equation}
        \epsilon\lambda^2\ll1\qquad(\textrm{relativistic regime}).
    \end{equation}

\begin{figure}[t!]
    \centering
    \includegraphics{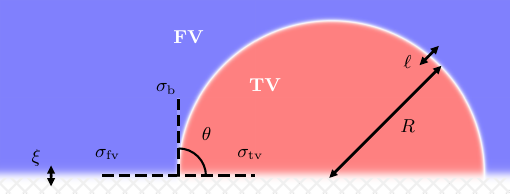}
    \caption{\label{fig:contact-angle}
    Edge nucleation in the thin-wall limit.
    The balance of surface tensions determines the contact angle $\theta$ between the bubble wall and the boundary.
    Here we show a simple `bucket' trap, with zero density at the boundary.
    In this case we find $\theta=\uppi/2$, so that edge nucleation forms half a bubble.
    For traps with a high-density `trench' we instead find $\theta=\uppi$, so it is only possible to form whole bubbles in the bulk.
    Also shown are the critical bubble radius $R$, the bubble wall thickness $\ell\ll R$, and the healing length $\xi\ll\ell$, which sets the scale over which the atomic density `heals' to its bulk value from the boundary.}
\end{figure}

For $\lambda>1$, the potential~\eqref{eq:phi-potential} contains metastable local minima at $\phi=\uppi\pmod{2\uppi}$ that can undergo vacuum decay, spontaneously nucleating bubbles of `true vacuum' in which $\phi=0\pmod{2\uppi}$.
These nucleation events can be described in terms of a `bounce' solution $\phi_\mathrm{b}(\tau,\vb*x)$ to Eq.~\eqref{eq:eom} in Euclidean time $\tau=\rmi t$.
Here the crucial quantity is the Euclidean action of this solution,
    \begin{equation}
    \label{eq:euclidean-action}
        S=\!\int\!\dd{\tau}\dd[d]{\vb*x}\bigg[\frac{1}{2}\frac{{(\partial_\tau\phi_\mathrm{b})}^2}{c_\phi^2}+\frac{1}{2}\abs{\grad\phi_\mathrm{b}}^2+U(\phi_\mathrm{b})-U(\phi_\mathrm{fv})\bigg],
    \end{equation}
    which sets the nucleation rate, $\Gamma\sim\exp(-S/\hbar)$.
In the limit where the bubble wall (the region over which $\phi$ interpolates between true and false vacua) is much thinner than the radius of the bubble, the action can be written as $S=A\,\sigma_\mathrm{b}-V\upDelta U$, where $A$ and $V$ are the bubble's $(d+1)$-dimensional Euclidean surface area and volume, $\sigma_\mathrm{b}$ is the surface tension of the bubble wall, and $\upDelta U=U(\phi_\mathrm{fv})-U(\phi_\mathrm{tv})$ is the excess energy density in the false vacuum.
This thin-wall approximation is valid when the potential barrier between vacua is large, $\lambda\gg1$, while staying in the relativistic regime $\epsilon\lambda^2\ll1$.
This implies a hierarchy between the critical bubble radius $R$, the bubble wall thickness $\ell$, and the scale associated with density gradients in the condensate $\xi$ (the `healing length'),
    \begin{equation}
    \label{eq:scale-hierarchy}
    \begin{split}
        R&\sim\lambda\ell\gg\ell\\
        \ell&\sim\xi/\epsilon^{1/2}\gg\xi
    \end{split}
    \qquad
    \begin{split}
        &(\textrm{thin-wall regime}),\\
        &(\textrm{relativistic regime}).
    \end{split}
    \end{equation}
These three lengthscales are illustrated in Fig.~\ref{fig:contact-angle}.
In practice $\lambda$ will likely be not much larger than unity, as this enhances the decay rate, increasing the probability of seeing bubbles in a given experimental run.
However, our thin-wall results below still give useful insights into edge nucleation.
We confirm these insights numerically in Sec.~\ref{sec:sims}.

%-----------------------------------------------------------------------------%
\subsection{Edge nucleation}
\begin{figure}[b!]
    \centering
    \includegraphics{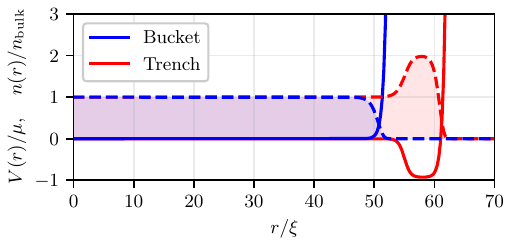}
    \caption{\label{fig:trapping-potentials}
    Trapping potentials used in our simulations, in units of the chemical potential $\mu=\hbar^2/(2m\xi^2)$ (solid curves), and corresponding profiles for the mean density $n=(n_\down+n_\up)/2$ (dashed curves).
    Both depend only on the radial coordinate $r$, measured in units of the healing length.
    Blue curves show a bucket trap, with a sharp wall at $r\approx50\,\xi$.
    Red curves show a trap with a trench layer, in which the density reaches double its bulk value.}
\end{figure}

The discussion above describes bulk nucleation far from the system's boundary; we now consider nucleation on the boundary, assuming a `bucket' potential of the kind shown in Fig.~\ref{fig:trapping-potentials}.
The number densities $n_\down$, $n_\up$ vanish at the boundary, and `heal' back to their bulk values over a lengthscale $\xi$.
The resulting density profile is sensitive to $\phi$, due to the associated energy density~\eqref{eq:phi-potential}.
There are thus three interfaces, each with its own surface tension: the false vacuum--boundary interface, with tension $\sigma_\mathrm{fv}$; the true vacuum--boundary interface, with tension $\sigma_\mathrm{tv}$; and the false vacuum--true vacuum interface (the bubble wall), with tension $\sigma_\mathrm{b}$.
These are illustrated in Fig.~\ref{fig:contact-angle}.
By resolving forces at the point where these surfaces meet, we find that the contact angle $\theta$ between the bubble wall and the boundary obeys
    \begin{equation}
    \label{eq:contact-angle}
        \cos\theta=\frac{\sigma_\mathrm{fv}-\sigma_\mathrm{tv}}{\sigma_\mathrm{b}}.
    \end{equation}
This is a well-known result in fluid mechanics, where it is known as Young's law~\cite{Kashchiev:2000nuc,Gallo:2021het,Caneletti:2024kww}.

Since we have a microphysical description of the system, one can calculate each of the three surface tensions in Eq.~\eqref{eq:contact-angle} to determine the contact angle analytically.
We describe this calculation in App.~\ref{sec:tensions}.
The key finding is that $\sigma_\mathrm{fv}-\sigma_\mathrm{tv}=\order{\epsilon}$, while $\sigma_\mathrm{b}=\order*{\epsilon^{1/2}}$.
Heuristically, this is because the relevant energy densities are $\order{\epsilon}$, but the bubble wall is thicker than the healing length by a factor $\sim\epsilon^{-1/2}$, cf. Eq.~\eqref{eq:scale-hierarchy}.
Since we require $\epsilon\ll1$ for a relativistic analog, Eq.~\eqref{eq:contact-angle} implies $\theta\simeq\uppi/2$, i.e., the bubble wall must be perpendicular to the boundary at the point of contact.
For a planar boundary, this means that edge nucleation forms half a bubble, as indicated in Fig.~\ref{fig:contact-angle}.

This is potentially a serious problem for analog vacuum decay, as can be appreciated by considering the Euclidean action~\eqref{eq:euclidean-action}.
Since the volume and surface area of an edge bubble are halved compared to a bulk bubble, so is its action: $S_\mathrm{edge}\simeq\frac{1}{2}S_\mathrm{bulk}$.
(In principle one should include an additional term to account for the excess tension at the true vacuum--boundary interface, $\sigma_\mathrm{tv}-\sigma_\mathrm{fv}$, but since this is a factor $\sim\epsilon^{1/2}$ smaller than $\sigma_\mathrm{b}$ it has negligible effect on the decay rate.)
Edge nucleation is therefore much faster than bulk nucleation, due to the exponential sensitivity of the decay rate to the Euclidean action.

An immediate consequence of this is that any corners in the box trap will cause even faster nucleation by forming an even smaller fraction of a bulk bubble (e.g., a quarter of a bubble for a right-angled corner in a 2D system), as seen in the results of Ref.~\cite{Billam:2022ykl}.
We therefore consider only circular traps in our simulations in Sec.~\ref{sec:sims}.
Circular symmetry is also convenient for numerical reasons, as discussed in App.~\ref{sec:trap-ics}.

\begin{figure*}[t!]
    \centering
    \includegraphics{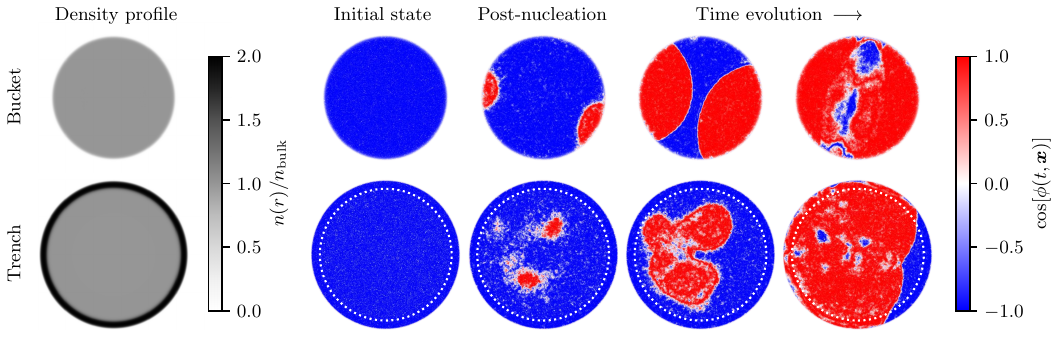}
    \caption{\label{fig:sims}
    Representative simulation results.
    The top and bottom rows show `bucket' and `trench' potentials, respectively, with white dotted circles in the bottom row indicating the inner edge of the trench.
    In the bucket, the false vacuum (blue) decays by nucleating edge bubbles (red), which meet the boundary at an angle $\theta\simeq\uppi/2$.
    In the trench, edge nucleation is prevented, and bulk nucleation dominates instead.
    The resulting bubbles are more noticeably aspherical due to the different parameters being simulated, corresponding to a shallower false vacuum.}
\end{figure*}

%-----------------------------------------------------------------------------%
\subsection{Eliminating edge nucleation}
The solution to the problem identified above is to modify the contact angle $\theta$ by engineering the trapping potential.
This could be implemented in practice using, e.g., a digital micromirror device to imprint the desired optical potential~\cite{Gauthier:2016dmd,Zou:2021opt}.
In particular, setting $\theta=\uppi$ immediately solves the edge nucleation problem; bubbles can then graze the boundary but not intersect it.
Bulk nucleation of spherical bubbles then becomes the minimum-action Euclidean solution and the dominant decay channel, as it is in the early-Universe scenarios we wish to probe.

This can be achieved by creating a `trench' layer inside the walls of the box, in which the potential is lower than in the bulk and the mean density $n=(n_\down+n_\up)/2$ is higher (see Fig.~\ref{fig:trapping-potentials}).
The Euclidean action~\eqref{eq:euclidean-action} is directly proportional to $n$ due to the self-interaction energy of the condensate, so all nucleation processes in the trench are exponentially slower than those in the bulk.
We can therefore treat the trench as being trapped in the false vacuum on the timescales relevant to bulk nucleation.
Crucially, this means that the interface between a true vacuum bubble in the bulk and the high-density false vacuum in the trench consists of both a phase profile (a bubble wall) \emph{and} a density profile.
As we show in detail in App.~\ref{sec:tensions}, this means that the associated tension is just the sum of the bubble wall tension and the false vacuum bulk--trench interface tension, $\sigma_\mathrm{tv}=\sigma_\mathrm{fv}+\sigma_\mathrm{b}$.
Comparing with Eq.~\eqref{eq:contact-angle}, we see that the contact angle is $\theta=\uppi$ as desired; the energy cost of an interface with the trench repels any bubbles, and bulk nucleation of spherical bubbles is preferred, solving the edge nucleation problem.

%%%%%%%%%%%%%%%%%%%%%%%%%%%%%%%%%%%%%%%%%%%%%%%%%%%%%%%%%%%%%%%%%%%%%%%%%%%%%%%
\section{Lattice simulations}\label{sec:sims}
The arguments in Sec.~\ref{sec:edges} provide evidence for the edge nucleation problem, and for a solution in the form of a trench potential.
Here we verify these predictions using real-time lattice simulations.
There are two reasons for doing this.
First, the analytical predictions rely on a thin-wall approximation, valid in the limit $\lambda\gg1$, whereas much of the experimentally-accessible parameter space is in the thick-wall regime, $\lambda\sim1$.
Second, these predictions rely on the Euclidean instanton formalism, whereas one of the core goals of the analog vacuum decay program is to test this formalism.
In particular, our goal is to test whether these idealized imaginary-time predictions are borne out in the real-time evolution of the system.

We use semiclassical simulations, in which the initial state contains random draws of the vacuum fluctuations in the $\psi_\down$, $\psi_\up$ fields, which are evolved forward in real time using the classical equations of motion~\cite{Braden:2018tky}.
By running a large ensemble of such simulations, one can approximate quantum expectation values of observables with ensemble averages.
This approach underpins many numerical simulations of early-Universe phenomena~\cite{Khlebnikov:1996mc,Rajantie:2000nj,Garcia-Bellido:2007nns,Amin:2011hj,Clough:2016ymm}, and is ubiquitous in atomic physics and quantum optics, where it is known as the truncated Wigner approximation~\cite{Braden:2018tky,Blakie:2008vka}.
In the context of vacuum decay, these simulations complement the Euclidean formalism by providing an alternative description that is valid to the same semiclassical order, but gives much richer dynamical information about the system.
While many of the predictions of the Euclidean formalism have been reproduced using these simulations, they tend to predict significantly faster nucleation rates~\cite{Braden:2018tky,Jenkins:2023eez} (though accounting for renormalization effects might resolve this discrepancy~\cite{Braden:2022odm}).
Analog experiments will eventually shed light on the relationship between these approaches and how well they approximate the quantum dynamics.
For our purposes here, the lattice simulations are simply a cross-check of the predictions in Sec.~\ref{sec:edges}.
We find that the two approaches are in complete agreement on the question of edge nucleation.

%-----------------------------------------------------------------------------%
\subsection{Physical parameters}
We simulate a quasi-2D system in which the atoms are tightly vertically confined by a harmonic potential $V_\bot(z)=\tfrac{1}{2}m\omega_\bot^2z^2$.
This is preferable to a 3D system as it allows the entire field to be directly imaged (rather than being reconstructed from line-of-sight-integrated images), and avoids challenges associated with levitating both atomic species equally against gravity.
Our states $\ket{\down},\ket{\up}$ are the $\ket{F,m_F}=\ket{1,0},\ket{1,-1}$ hyperfine states of ${}^{39}\mathrm{K}$.
In a uniform magnetic field $B\approx57.5\,\mathrm{G}$ the two-body interactions between these states are such that the relativistic equation of motion~\eqref{eq:eom} is achieved by setting $(n_\down-n_\up)/(n_\down+n_\up)\approx0.298$~\cite{Lysebo:2010fesh,Jenkins:2023npg,Brown:2025wxy}.

The nucleation rates for bulk and edge bubbles are both set by the dimensionless density $N\xi^2/A$, where $N$ is the total atom number, $A$ is the 2D volume of the system, and $\xi$ is the healing length.
One can tune this parameter while keeping $\xi$ fixed by varying the transverse trapping frequency as $\omega_\bot\propto N^{-1/2}$~\cite{Jenkins:2023eez}.
We consider two cases: a `high density' setup with $N=4.80\times10^5$, $\omega_\bot=32.9\times2\uppi\,\mathrm{kHz}$, and a `low density' setup with $N=2.40\times10^5$, $\omega_\bot=132\times2\uppi\,\mathrm{kHz}$.
In both cases we consider a circular 2D trap with radius $r\approx50\,\xi$.
We set $\xi=1\,\upmu\mathrm{m}$, which is typical of quasi-2D cold-atom experiments~\cite{Karailiev:2024sxs,Viermann:2022wgw}.
For the Rabi coupling we take $\epsilon=4.11\times10^{-3}$, corresponding to a mean Rabi frequency $\Omega_0=18.8\times2\uppi\,\mathrm{Hz}$, and $\lambda=1.1$, corresponding to a modulation amplitude $\upDelta\Omega=9.97\times10^{-2}\,\nu$, with $\nu\gtrsim\mathrm{MHz}$ the modulation frequency.

We simulate this setup using a Fourier pseudospectral lattice code with an eighth-order symplectic time-stepping scheme---see Refs.~\cite{Jenkins:2023eez,Jenkins:2023npg} for details.
We use a $1024\times1024$ periodic square lattice with spacing $\updelta x=0.190\,\xi$ and timestep $\updelta t=0.0362\,\hbar/(m_\phi^{}c_\phi^2)$; this allows a gap of $\approx42.5\,\xi$ between the box trap walls and each end of the lattice, which is large enough that the system is insensitive to the periodicity.
We run each simulation up to time $t=1180\,\hbar/(m_\phi^{}c_\phi^2)$, which is roughly double the sound-crossing time across the condensate.

We simulate two axisymmetric trapping potentials of the form
    \begin{equation}
    \begin{split}
        V(r)&=\frac{1}{2}V_{\max}\qty[1+\tanh(\frac{r-r_0-w}{\xi})]\\
        &+\frac{1}{2}V_\mathrm{trench}\qty[\tanh(\frac{r-r_0-w}{\xi})-\tanh(\frac{r-r_0}{\xi})],
    \end{split}
    \end{equation}
    where $V_{\max}=841\,\hbar^2/(m\xi^2)$ is the height of the potential barrier (which is large to prevent high-momentum modes escaping) and $r_0=55.0\,\xi$ is the approximate radial size of the bulk region.
For $V_\mathrm{trench}=w=0$ we recover a `bucket' potential; we also simulate a potential with a trench of depth $V_\mathrm{trench}=\hbar^2/(2m\xi^2)$ and width $w=9.75\,\xi$.
These potentials are shown in Fig.~\ref{fig:trapping-potentials}, along with the corresponding condensate density profiles, which we compute by evolving the equations of motion for $\psi_\down$, $\psi_\up$ in imaginary time from homogeneous initial conditions~\cite{Choi:1998ia}.

\begin{figure}[t!]
    \centering
    \includegraphics{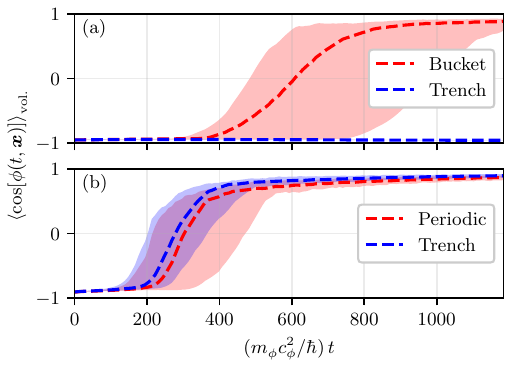}
    \caption{\label{fig:cos-mean}
    Time evolution of the volume-averaged cosine of the relative phase for each ensemble of simulations.
    This quantity serves as a vacuum decay diagnostic, starting near the false vacuum value $\cos\phi_\mathrm{fv}=-1$, and transitioning toward the true vacuum value $\cos\phi_\mathrm{tv}=+1$ after bubble nucleation (whether on the edge or in the bulk).
    Dashed curves show median values as functions of time, while shaded regions contain $95.5\%$ probability ($\pm2\sigma$ Gaussian equivalent).}
\end{figure}

%-----------------------------------------------------------------------------%
\subsection{Results}
We investigate bubble nucleation in these two potentials by running ensembles of 512 simulations.
Each simulation has an independent random realization of the vacuum fluctuations around the background condensate, generated by populating the energy eigenmodes above the false vacuum; see App.~\ref{sec:trap-ics} for details on how we compute these modes for each trapping potential.
We carry out two comparisons between ensembles to test the predictions of Sec.~\ref{sec:edges}.

First, we compare nucleation in the bucket and trench traps for the `high density' parameters described above.
We choose these parameters such that the bulk nucleation timescale is much longer than the simulation time, so that no bubbles should form in the absence of boundary effects.
However, the bucket trap causes the system to decay well within the simulation time, as shown in Fig.~\hyperref[fig:cos-mean]{5(a)}.
We confirm that, as predicted in Sec.~\ref{sec:edges}, every simulation in the bucket-trap ensemble decays by nucleating one or more edge bubbles, which meet the boundary at an angle $\theta\simeq\uppi/2$ (see Fig.~\ref{fig:sims}).
Modifying the trapping potential completely eliminates this decay channel, with every simulation in the trench-trap ensemble surviving to the end of the simulation time.

Second, we compare nucleation in the trench trap and in a periodic system with no trap for the `low density' parameters described above.
The periodic simulations are carried out on a $512\times512$ lattice with the same spacing $\updelta x$, such that the periodic volume is approximately equal to that of the trap interior.
We choose the `low density' parameters so that the Euclidean action~\eqref{eq:euclidean-action} associated with bulk nucleation is approximately half of that in the `high density' case; we therefore expect bulk nucleation in the low-density setup to occur at a comparable rate to edge nucleation in the high-density setup.
As shown in Fig.~\hyperref[fig:cos-mean]{5(b)}, this is indeed the case: both the trench and periodic ensembles decay at essentially the same rate.
This confirms that the rate is insensitive to the boundary once the trench has been implemented.
We also confirm that every simulation in this low-density trench ensemble decays via bulk nucleation (see bottom row of Fig.~\ref{fig:sims}).

The bubbles that form in the low-density simulations are noticeably more distorted and aspherical than those in the high-density simulations; this is an expected consequence of having higher-amplitude vacuum fluctuations, which renormalize the effective potential~\eqref{eq:phi-potential}, resulting in a shallower false vacuum potential barrier~\cite{Braden:2022odm} and therefore thicker bubble walls and faster decay rates.
In the limit where the barrier vanishes, the system undergoes global spinodal decomposition rather than forming localized bubbles.
Here we are still in the regime of well-defined bubble nucleation events, but nonetheless see significant deviations from the standard paradigm of extremely rare and highly spherical thin-wall bubbles~\cite{Coleman:1977py,Callan:1977pt}.
This highlights an advantage of the analog experiments and real-time lattice simulations: both allow one to study relativistic bubble nucleation in regimes where the Euclidean description breaks down.
In fact, practical limitations on experimental coherence times and numerical runtimes mean that thick-wall bubbles are the easiest to access with these methods; the long timescales associated with spherical thin-wall bubbles make them more challenging to access in 2D simulations, and potentially also in the experiments.
Theoretical uncertainties in the nucleation rate~\cite{Braden:2018tky,Braden:2022odm} make it difficult to quantify exactly how far 2D experiments can reach into the thin-wall regime within realistic coherence times (typically $\sim1\,\mathrm{s}~\cite{Pethick:2008bec}$).
However, this issue can always be circumvented using a 1D setup, for which the decay rate is parametrically faster~\cite{Jenkins:2023eez,Jenkins:2023npg}.
Future experiments will therefore be able to probe relativistic bubble nucleation across these different regimes, yielding insights into a broad range of cosmological scenarios.

%-----------------------------------------------------------------------------%
\subsection{Experimental considerations}
While we have focused on one specific form of the trench-trap potential in our simulations, the theoretical understanding developed in Sec.~\ref{sec:edges} and App.~\ref{sec:tensions} suggests that our findings should extend to a very broad family of such potentials.
There are only two essential requirements: first, the trench should be deep enough that the atomic density is significantly higher there than in the bulk; and second, it should be wide enough that the condensate is able to attain this enhanced density by `healing' from the bulk value, before being damped to zero at the edge of the system.
Increasing the density in the trench exponentially suppresses any nucleation processes that would otherwise occur there, and thus also makes the setup more robust against imperfections such as noise in the optical potential.

Another consideration is the spatial resolution of the trench.
Previous experiments have demonstrated sub-micron resolution of the optical potential imprinted using digital micromirror devices~\cite{Gauthier:2016dmd,Zou:2021opt}, with arbitrary control of the strength of the potential achieved by averaging many pixels within this sub-micron diffraction pattern.
Crucially, this resolution is smaller than the typical healing length $\xi\sim1\,\mu\mathrm{m}$, which limits the scale over which the condensate density responds to changes in the potential.
Any finer resolution is therefore unneccesary, as it would have little effect on the resulting density profile.

It will likely be desirable to make the trench as shallow and as narrow as possible while still successfully eliminating edge nucleation, to minimize the fraction of condensed atoms that are required to populate the trench.
The flexible and generic nature of the trench-trap solution suggest that it should be possible to achieve this trade-off without excessive fine-tuning of the potential.

%%%%%%%%%%%%%%%%%%%%%%%%%%%%%%%%%%%%%%%%%%%%%%%%%%%%%%%%%%%%%%%%%%%%%%%%%%%%%%%
\section{Summary and outlook}\label{sec:summary}
Cold-atom analog experiments will soon enable empirical tests of relativistic false vacuum decay in the laboratory, giving new insights into the physics of the very early Universe.
A key challenge for this program is ensuring the faithfulness of the early-Universe analogy by characterizing and mitigating any noncosmological behavior.
In this paper we have identified the presence of boundaries in the system as potentially problematic for analog vacuum decay, showing that they generically lead to rapid decay via nucleation of `edge bubbles' that have no cosmological counterpart.
However, we have shown that this problem can be straightforwardly eliminated by engineering the optical potential used to trap the atoms: creating a high-density `trench' layer prohibits edge nucleation, and allows one to observe the bulk nucleation that is relevant for cosmology.
Identical conclusions are found in a companion paper~\cite{Brown:2025wxy}, which investigates thermal nucleation in a broader range of analog systems.
This trench solution demonstrates how current experimental capabilities---e.g., the ability to imprint highly customizable optical traps using digital micromirror devices~\cite{Gauthier:2016dmd,Zou:2021opt}---enable faithful analog simulations of early-Universe theories.

Our results bring us a step closer to simulating vacuum decay with cold atoms.
There remain further experimental complications that we plan to investigate in future work.
These include characterizing the effects of various noise sources (including magnetic field noise and fluctuations in the optical potential), as well as better understanding the small-scale behavior of the system, particularly regarding the damping of instabilities associated with the modulated Rabi coupling~\cite{Braden:2019vsw}.
There are also important open questions regarding how the effective potential~\eqref{eq:phi-potential} is renormalized by small-scale modes, and how these corrections differ from the pure Klein-Gordon case studied in Ref.~\cite{Braden:2022odm}.
Understanding these issues will afford us greater control over our theoretical predictions, allowing us to extract the maximum possible insight into cosmological physics from upcoming experiments.

%%%%%%%%%%%%%%%%%%%%%%%%%%%%%%%%%%%%%%%%%%%%%%%%%%%%%%%%%%%%%%%%%%%%%%%%%%%%%%%
\begin{acknowledgments}
    We are grateful to Zoran Hadzibabic for insights and suggestions that made this work possible.
    We thank Tom Billam, Kate Brown, Christoph Eigen, Emilie Hertig, Matt Johnson, Konstantinos Konstantinou, Ian Moss, Tanish Satoor, Feiyang Wang, Silke Weinfurtner, Paul Wong, and Yansheng Zhang for fruitful discussions.

    This work was supported by the Science and Technology Facilities Council (STFC) through the UKRI Quantum Technologies for Fundamental Physics Programme (grant number ST/T005904/1).
    ACJ was supported by the Engineering and Physical Sciences Research Council (EPSRC) through a Stephen Hawking Fellowship (grant number EP/U536684/1), and by the Gavin Boyle Fellowship at the Kavli Institute for Cosmology, Cambridge.

    Part of this work was carried out at the Munich Institute for Astro-, Particle and BioPhysics (MIAPbP), which is funded by the Deutsche Forschungsgemeinschaft (DFG, German Research Foundation) under Germany's Excellence Strategy – EXC-2094 – 390783311.
    ACJ is grateful for hospitality at Newcastle University, where part of this work was carried out.

    Our simulations were performed on the Hypatia cluster at UCL, using computing equipment funded by the Research Capital Investment Fund (RCIF) provided by UKRI, and partially funded by the UCL Cosmoparticle Initiative.
    We are grateful to Edd Edmondson for technical support.
    We acknowledge the use of the Python packages NumPy~\cite{Harris:2020xlr}, SciPy~\cite{Virtanen:2019joe}, and Matplotlib~\cite{Hunter:2007ouj}.

    The data that support the findings of this study are openly available~\cite{2504.02829-data}.
\end{acknowledgments}

%%%%%%%%%%%%%%%%%%%%%%%%%%%%%%%%%%%%%%%%%%%%%%%%%%%%%%%%%%%%%%%%%%%%%%%%%%%%%%%
\section*{Author contributions}

Contributions based on the CRediT (Contributor Role Taxonomy) system.
\textbf{ACJ}: conceptualization; methodology; software; formal analysis; investigation; data curation; interpretation and validation;  visualization; writing (original draft).
\textbf{HVP}: conceptualization; interpretation and validation; writing (review); project administration.
\textbf{AP}: conceptualization; interpretation and validation; writing (review).

%%%%%%%%%%%%%%%%%%%%%%%%%%%%%%%%%%%%%%%%%%%%%%%%%%%%%%%%%%%%%%%%%%%%%%%%%%%%%%%
\appendix
\section{Surface tension calculations}\label{sec:tensions}
In this Appendix we calculate the surface tensions associated with each of the interfaces shown in Fig.~\ref{fig:contact-angle}, with the goal of deriving contact angles of $\theta\simeq\uppi/2$ and $\theta=\uppi$ in the case of the bucket and trench traps, respectively.

The cold-atom system is described by the Hamiltonian density~\cite{Jenkins:2023npg}
    \begin{equation}
    \label{eq:hamiltonian}
    \begin{split}
        \mathcal{H}&=-\psi_\down^\dagger\frac{\hbar^2\laplacian}{2m}\psi_\down-\psi_\up^\dagger\frac{\hbar^2\laplacian}{2m}\psi_\up+(V\!-\!\mu)(\psi_\down^\dagger\psi_\down+\psi_\up^\dagger\psi_\up)\\
        &-\frac{\hbar\Omega}{2}(\psi_\down^\dagger\psi_\up+\psi_\up^\dagger\psi_\down)-\frac{\hbar\delta}{2}(\psi_\down^\dagger\psi_\down-\psi_\up^\dagger\psi_\up)\\
        &+\sum_{i,j=\down,\up}\frac{g_{ij}}{2}\psi^\dagger_i\psi^\dagger_j\psi_i\psi_j,
    \end{split}
    \end{equation}
    where $V(\vb*x)$ is the trapping potential, $\mu$ is the chemical potential, and $g_{ij}$ is the effective 2D two-body interaction between atomic states $\ket{i}$ and $\ket{j}$; these interactions are more conveniently described in terms of the linear combinations
    \begin{equation}
        g=\frac{g_{\down\down}+g_{\up\up}}{2},\quad\Delta=\frac{g_{\up\up}-g_{\down\down}}{2},\quad\kappa=\frac{g_{\down\down}+g_{\up\up}-2g_{\down\up}}{2}.
    \end{equation}
The Rabi coupling is described by the two terms on the second line of Eq.~\eqref{eq:hamiltonian}; $\Omega(t)$ is the Rabi frequency, set by the (time-varying) amplitude of the applied radio-frequency field, while $\delta$ is the detuning (i.e., the difference between the frequency of the applied field and the resonant frequency for the two-state system).

The relativistic analog is obtained using a Rabi frequency with a small constant piece (which generates the cosine term in the potential~\eqref{eq:phi-potential}) and a rapidly modulated piece (which generates the false vacuum potential barrier),
    \begin{equation}
        \Omega(t)=2\epsilon\sqrt{\kappa^2-\Delta^2}\,\frac{\bar{n}_\mathrm{fv}}{\hbar}+\sqrt{2\epsilon}\,\lambda\nu\cos\nu t,
    \end{equation}
    where $\epsilon\ll1$ and $\lambda\ge1$ are the dimensionless parameters introduced in Eq.~\eqref{eq:phi-potential}, $\nu$ is the modulation frequency, which is much larger than all other frequencies in the system, and $\bar{n}_\mathrm{fv}$ is the mean number density per species in the uniform false vacuum state.
(The equilibrium number density depends on the energy density of the system and therefore on the value of the relative phase, hence the distinction in defining $\bar{n}_\mathrm{fv}$.)
The detuning is assumed to be small, $\delta=\order{\epsilon}$.

On timescales longer than the modulation period $2\uppi/\nu$, the system is well-described by the effective Hamiltonian~\cite{Jenkins:2023npg}
    \begin{equation}
    \label{eq:hamiltonian-time-averaged}
    \begin{split}
        \mathcal{H}_\mathrm{eff}&=-\psi_\down^\dagger\frac{\hbar^2\laplacian}{2m}\psi_\down-\psi_\up^\dagger\frac{\hbar^2\laplacian}{2m}\psi_\up\\
        &+(V-\mu)(\psi_\down^\dagger\psi_\down+\psi_\up^\dagger\psi_\up)\\
        &-\epsilon\bar{n}_\mathrm{fv}\sqrt{\kappa^2-\Delta^2}(\psi_\down^\dagger\psi_\up+\psi_\up^\dagger\psi_\down)-\frac{\hbar\delta}{2}(\psi_\down^\dagger\psi_\down-\psi_\up^\dagger\psi_\up)\\
        &+\frac{1}{2}\qty(g-\Delta-\frac{\epsilon\lambda^2}{2}(\kappa-\Delta))\psi_\down^\dagger\psi_\down^\dagger\psi_\down\psi_\down\\
        &+\frac{1}{2}\qty(g+\Delta-\frac{\epsilon\lambda^2}{2}(\kappa+\Delta))\psi_\up^\dagger\psi_\up^\dagger\psi_\up\psi_\up\\
        &+(g-\kappa(1-\epsilon\lambda^2))\psi_\down^\dagger\psi_\down\psi_\up^\dagger\psi_\up\\
        &-\frac{\epsilon\lambda^2}{4}\kappa(\psi_\down^\dagger\psi_\down^\dagger\psi_\up\psi_\up+\psi_\up^\dagger\psi_\up^\dagger\psi_\down\psi_\down)+\order{\epsilon^2},
    \end{split}
    \end{equation}
    where the terms proportional to $\lambda^2$ are generated by `integrating out' the rapid modulation of the Rabi coupling.
This is done perturbatively in the small parameter $\epsilon$ associated with the mean value of the Rabi coupling; from now on we implicitly neglect all terms of order $\epsilon^2$.
We also neglect the `eff' subscript below, as all of our calculations are performed using this effective Hamiltonian.

%-----------------------------------------------------------------------------%
\subsection{Constraint equations for static solutions}
We look for static solutions to the equations of motion generated by the effective Hamiltonian~\eqref{eq:hamiltonian-time-averaged}, allowing us to describe the various interfaces shown in Fig.~\ref{fig:contact-angle} at rest, as appropriate at the moment of bubble nucleation.
We therefore set
    \begin{equation}
    \label{eq:static-gpe}
        \rmi\hbar\partial_t\psi_\down=\pdv{\mathcal{H}}{\psi_\down^\dagger}=0,\qquad\rmi\hbar\partial_t\psi_\up=\pdv{\mathcal{H}}{\psi_\up^\dagger}=0.
    \end{equation}
It is convenient to write the atomic fields as
    \begin{equation}
    \label{eq:psi-ansatz}
    \begin{split}
        \psi_\down&=\sqrt{n(1+z)}\exp(\frac{\epsilon z}{2(1+z)}\chi+\frac{1-z}{2}\rmi\phi),\\
        \psi_\up&=\sqrt{n(1-z)}\exp(-\frac{\epsilon z}{2(1-z)}\chi-\frac{1+z}{2}\rmi\phi),
    \end{split}
    \end{equation}
    where $n(\vb*x)$ is the mean number density per species and $\phi(\vb*x)$ is the relative phase, which admits an effective relativistic description on large scales.
The background population imbalance $z$ is treated as a spatially uniform constant, as is the total phase degree of freedom $\theta=(1+z)\phi_\down+(1-z)\phi_\up$, which we set to zero everywhere.
These choices are self-consistent for a critical value of $z$ (which we derive below), for which the relative and total phase fields decouple from each other~\cite{Jenkins:2023npg}.
Finally, the field $\chi(\vb*x)$ generates $\order{\epsilon}$ perturbations in the population imbalance (associated with variations in $\phi$) which leave $n$ unchanged.
We assume that spatial derivatives of $\chi$ are suppressed by further powers of $\epsilon$ and can thus be neglected; this is because both $\chi$ and $\phi$ vary on lengthscales that are parametrically larger than the healing length in the relativistic regime we are interested in (cf.~Eq.~\eqref{eq:scale-hierarchy}).
This assumption, and the ansatz~\eqref{eq:psi-ansatz} more generally, are justified \emph{a posteriori} by the self-consistent solutions we find below.

Inserting this ansatz into Eq.~\eqref{eq:static-gpe}, we set the real and imaginary parts of both equations to zero to obtain a set of constraint equations that characterize static configurations of the system.
Naively there are four such equations (two from each complex field $\psi_\down$, $\psi_\up$), but eliminating the total phase removes one of these, leaving three constraints.
In the simplest case of a homogeneous false vacuum state with $n=\bar{n}_\mathrm{fv}$, the potential $V$ and all spatial derivatives vanish, leaving a set of algebraic equations that is solved by fixing the chemical potential, background population imbalance, and Rabi detuning,
    \begin{equation}
    \label{eq:constraints-fv-background}
    \begin{split}
        \mu&=\qty(2g-\frac{\kappa^2+\Delta^2}{\kappa}+\epsilon\kappa)\bar{n}_\mathrm{fv},\\
        z&=\frac{\Delta}{\kappa}\qty(1+\frac{\epsilon\lambda^2}{2}),\\
        \hbar\delta&=-2\epsilon\Delta\bar{n}_\mathrm{fv}.
    \end{split}
    \end{equation}
(For later convenience, we define $\mu_0=[2g-(\kappa^2+\Delta^2)/\kappa]\bar{n}_\mathrm{fv}$ and $z_0=\Delta/\kappa$ as the $\order{\epsilon^0}$ parts of the chemical potential and population imbalance.)
Since these three quantities are all constant, Eq.~\eqref{eq:constraints-fv-background} fixes their values for inhomogeneous and non-false-vacuum solutions too.
We therefore insert these values into Eq.~\eqref{eq:static-gpe} to obtain the general constraint equations,
    \begin{equation}
    \label{eq:constraint-equations}
    \begin{split}
        \frac{\hbar^2}{4m}\grad\vdot(n\grad\phi)&=\epsilon\kappa n\qty(\bar{n}_\mathrm{fv}\sin\phi+\frac{\lambda^2}{2}n\sin2\phi),\\
        \frac{\hbar^2}{4m}\abs{\grad\phi}^2&=\epsilon\kappa\qty[\bar{n}_\mathrm{fv}(1+\cos\phi)+n(\chi-\lambda^2\sin^2\phi)],\\
        \frac{\hbar^2}{2m}\frac{\laplacian\sqrt{n}}{\sqrt{n}}&=V+\mu_0\frac{n-\bar{n}_\mathrm{fv}}{\bar{n}_\mathrm{fv}}\\
        &\!\!\!\!\!\!\!\!\!\!\!\!\!\!\!\!-\frac{\epsilon\kappa}{2}(1-z_0^2)\qty[\bar{n}_\mathrm{fv}(1+\cos\phi)-n(\chi+\lambda^2\sin^2\phi)].
    \end{split}
    \end{equation}

By solving these equations, we can evaluate the Hamiltonian density~\eqref{eq:hamiltonian-time-averaged} and thereby calculate the surface tension of each interface,
    \begin{equation}
    \label{eq:surface-tension}
        \sigma=\int_\mathcal{C}\dd{x}(\mathcal{H}-\bar{\mathcal{H}}_\mathrm{fv}),
    \end{equation}
    where the integral is along a curve $\mathcal{C}$ that passes through the interface and is orthogonal to it; we let $x$ denote a coordinate direction that is locally parallel to this curve.
$\mathcal{H}(x)$ is the energy density evaluated along $\mathcal{C}$, and $\bar{\mathcal{H}}_\mathrm{fv}$ is the constant background energy density associated with the homogeneous false vacuum.
Simple `on-shell' expressions for these can be found by substituting the constraint equations~\eqref{eq:constraint-equations} back into Eq.~\eqref{eq:hamiltonian-time-averaged},
    \begin{equation}
    \begin{split}
        \mathcal{H}&=-\mu_0\frac{n^2}{\bar{n}_\mathrm{fv}}-\epsilon\lambda^2\kappa(1-z_0^2)n^2\sin^2\phi,\quad\bar{\mathcal{H}}_\mathrm{fv}=-\mu_0\bar{n}_\mathrm{fv}.
    \end{split}
    \end{equation}
We outline the surface tension calculation for each of the three key cases in turn below.

%-----------------------------------------------------------------------------%
\subsection{Bubble wall tension}
For a bubble wall in the bulk, we set $V=0$ and solve for the small density perturbations $n-\bar{n}_\mathrm{fv}$ and $\chi$ as functions of $\phi$.
The constraint equations~\eqref{eq:constraint-equations} then combine to give
    \begin{equation}
    \label{eq:bubble-wall-profile}
        \partial_x\phi=\sqrt{2[U(\phi)-U(\phi_\mathrm{fv})]},
    \end{equation}
    where $U(\phi)$ is the potential in Eq.~\eqref{eq:phi-potential}, with $m_\phi=2m/\sqrt{1-z_0^2}$ and $c_\phi^2=(\kappa\bar{n}_\mathrm{fv}/m)(1-z_0^2)$.
This agrees exactly with the usual equation describing the structure of relativistic bubble walls~\cite{Coleman:1977py}; this is to be expected, given that we are working in the relativistic regime of the analog system.
Solving for the wall profile and performing the integral in Eq.~\eqref{eq:surface-tension} we find that the bubble wall tension is
    \begin{equation}
    \label{eq:bubble-wall-tension}
        \sigma_\mathrm{b}=\sqrt{4\epsilon\,\frac{\hbar^2\kappa\bar{n}_\mathrm{fv}^3}{m}}(1-z_0^2)I(\lambda),
    \end{equation}
    where $I(\lambda)$ is a dimensionless integral depending only on the barrier height $\lambda$, which approaches $I(\lambda)\to\lambda$ in the thin-wall limit $\lambda\gg1$.
Solving Eq.~\eqref{eq:bubble-wall-profile} also gives us the characteristic lengthscale associated with the bubble wall profile (i.e., its thickness), which is $\ell\simeq\hbar/(\epsilon^{1/2}\lambda m_\phi c_\phi)$ in the thin-wall regime $\lambda\gg1$, as well as the corresponding bubble radius $R\sim\lambda\ell$.
The key finding for our purposes is that $\sigma_\mathrm{b}=\order{\epsilon^{1/2}}$; as discussed in Sec.~\ref{sec:edges}, this is because the excess energy density in the bubble wall is $\order{\epsilon}$, but the thickness of the wall is $\ell=\order{\epsilon^{-1/2}}$.

%-----------------------------------------------------------------------------%
\subsection{Boundary tension: bucket trap}
For interfaces at the edge of the bucket trap, we treat the potential as an infinite planar hard wall,
    \begin{equation}
        V(x)=
        \begin{cases}
            0 & \qif*x>0,\\
            \infty & \qif*x<0,
        \end{cases}\qquad(\mathrm{bucket}).
    \end{equation}
This approximation allows us to derive closed-form analytical expressions for the surface tensions; however, we expect our key findings to be insensitive to this choice, so long as the potential `switches on' over a lengthscale comparable to the healing length $\xi=\hbar/\sqrt{2m\mu_0}$.

First we consider the false-vacuum--boundary interface, setting $\phi=\uppi$ everywhere.
The first of the constraint equations~\eqref{eq:constraint-equations} is then trivially satisfied, the second gives $\chi=0$, and the third reduces to give
    \begin{equation}
        \frac{\hbar^2}{2m}\frac{\partial_x^2\sqrt{n}}{\sqrt{n}}=V+\mu_0\frac{n-\bar{n}_\mathrm{fv}}{\bar{n}_\mathrm{fv}}.
    \end{equation}
This equation is identical to that describing a single-component condensate~\cite{Pethick:2008bec}.
Solving this in the interior region $x>0$, subject to the boundary condition $n=0$ imposed by the bucket potential at $x=0$, gives half of the well-known `dark soliton' solution to the Gross-Pitaevskii equation~\cite{Pethick:2008bec}, joined continuously to the region of zero density beyond the wall,
    \begin{equation}
        n(x)=
        \begin{cases}
            \bar{n}_\mathrm{fv}\tanh[2](\frac{x}{\sqrt{2}\xi}) & \qif*x>0,\\
            0 & \qif*x<0.
        \end{cases}
    \end{equation}
The resulting surface tension is then
    \begin{equation}
        \sigma_\mathrm{fv}=\frac{4\sqrt{2}}{3}\mu_0\bar{n}_\mathrm{fv}\xi.
    \end{equation}

The situation is very similar at the true-vacuum--boundary interface, except that the bulk number density that the condensate `heals' to away from the wall is enhanced by an $\order{\epsilon}$ correction to offset the lower potential energy density $U(\phi)$.
(One can see this by inserting $\phi=0$ into Eq.~\eqref{eq:constraint-equations}, as the second equation then gives $n\chi=-2\bar{n}_\mathrm{fv}$, which generates an $\order{\epsilon}$ constant offset in the third equation.)
This translates into an $\order{\epsilon}$ difference between the two surface tensions,
    \begin{equation}
    \label{eq:bucket-tensions}
        \sigma_\mathrm{tv}=\sigma_\mathrm{fv}{\qty(\frac{\bar{n}_\mathrm{tv}}{\bar{n}_\mathrm{fv}})}^{\!2}=\sigma_\mathrm{fv}\qty[1+4\epsilon\frac{\kappa\bar{n}_\mathrm{fv}}{\mu_0}(1-z_0^2)],
    \end{equation}
    where we have defined the number density in the uniform true vacuum, $\bar{n}_\mathrm{tv}=\bar{n}_\mathrm{fv}[1+2\epsilon\kappa\bar{n}_\mathrm{fv}(1-z_0^2)/\mu_0]$.

Putting Eqs.~\eqref{eq:bubble-wall-tension} and~\eqref{eq:bucket-tensions} together, we find that the contact angle~\eqref{eq:contact-angle} between the bubble wall and the bucket trap in the thin-wall limit is given by
    \begin{equation}
        \cos\theta\simeq-\frac{4\sqrt{2}}{3}\frac{\xi}{\lambda^2\ell}\qquad(\mathrm{bucket}).
    \end{equation}
Since $\xi\ll\ell$ in the relativistic regime, this yields $\theta\simeq\uppi/2$ as claimed in Sec.~\ref{sec:edges}.

%-----------------------------------------------------------------------------%
\subsection{Boundary tension: trench trap}
In the trench case, we instead write the potential as
    \begin{equation}
        V(x)=
        \begin{cases}
            0 & \qif*x>0,\\
            -\mu_0v & \qif*x<0,
        \end{cases}\qquad(\mathrm{trench}),
    \end{equation}
    where $v>0$ is a dimensionless constant that parametrizes the depth of the trench.
Solving Eq.~\eqref{eq:static-gpe} deep inside the trench with $\phi=\uppi$ then gives $n=(1+v)\,\bar{n}_\mathrm{fv}$ in the false vacuum.
Near the trench boundary we once again find a solution which connects part of a dark soliton to a constant density solution on the other side of the interface,
    \begin{equation}
    \label{eq:trench-density-profile}
        n(x)=
        \begin{cases}
            \bar{n}_\mathrm{fv} & \qif*x>0,\\
            (1+v)\,\bar{n}_\mathrm{fv}\tanh[2]((1+v)\frac{x_0-x}{\sqrt{2}\xi}) & \qif*x<0,
        \end{cases}
    \end{equation}
    where $x_0$ is chosen to ensure continuity at $x=0$.
Crucially, the healing of the density occurs entirely inside the trench, i.e., in the $x<0$ region.

At the true-vacuum--trench interface there is a phase profile as well as a density profile, as $\phi$ interpolates between the false vacuum in the trench and the true vacuum in the bulk.
The energy cost of this phase profile scales with the local number density (cf. Eq.~\eqref{eq:bubble-wall-tension}), so the lowest-energy configuration is to have it occur entirely in the $x>0$ region, where it becomes identical to a bulk bubble wall.
Meanwhile, Eq.~\eqref{eq:trench-density-profile} remains a valid solution for $x\le0$, interpolating between the high-density trench and the edge of the low-density bulk.
The total surface tension at the true-vacuum--trench interface therefore cleanly separates into a density contribution from the $x<0$ region, which is equal to the false-vacuum--trench tension $\sigma_\mathrm{fv}$, and a phase contribution from the $x>0$ region, which is equal to the bubble wall tension $\sigma_\mathrm{b}$.
We therefore obtain $\sigma_\mathrm{tv}=\sigma_\mathrm{fv}+\sigma_\mathrm{b}$, so that
    \begin{equation}
        \cos\theta=-1\qquad(\mathrm{trench}),
    \end{equation}
    yielding $\theta=\uppi$ as claimed in Sec.~\ref{sec:edges}.

Note that in deriving this result we have placed no requirements on the depth $v$ or width $w$ of the trench, demonstrating the flexibility and generality of this approach to preventing edge nucleation.
Our only implicit assumptions are that $v$ is large enough to ensure that nucleation in the trench is strongly suppressed, and that $w$ is large enough that the condensate can reach the enhanced density $(1+v)\,\bar{n}_\mathrm{fv}$ before tapering to zero.
These conditions are met so long as $v$ is not much smaller than unity, and $w$ is at least a few times larger than the healing length.
Our simulations in Sec.~\ref{sec:sims} use $v=1$ and $w=9.75\,\xi$.

%%%%%%%%%%%%%%%%%%%%%%%%%%%%%%%%%%%%%%%%%%%%%%%%%%%%%%%%%%%%%%%%%%%%%%%%%%%%%%%
\section{Trapped initial conditions}\label{sec:trap-ics}
In this Appendix we describe our procedure for generating initial conditions for our truncated Wigner simulations, which approximate the initial false vacuum state of the system.
It is crucial that this is done accurately, as previous work has shown that misspecifying the initial conditions can dramatically alter the nucleation rate~\cite{Jenkins:2023eez}.

We consider small quantum fluctuations in the atomic fields,
    \begin{equation}
    \begin{split}
        \hat{\psi}_\down(\vb*x)&=\sqrt{n(r)(1+z)}+\updelta\hat{\psi}_\down(\vb*x),\\
        \hat{\psi}_\up(\vb*x)&=-\qty[\sqrt{n(r)(1-z)}+\updelta\hat{\psi}_\up(\vb*x)],
    \end{split}
    \end{equation}
    where the minus sign is due to the $\uppi$ relative phase associated with the false vacuum.
We set the population imbalance $z=(1+\epsilon\lambda^2/2)\Delta/\kappa$ so that the total and relative phase fields decouple~\cite{Jenkins:2023npg}, and numerically solve for the circularly-symmetric background number density $n(r)$ by evolving the equations of motion in imaginary time to find the ground state~\cite{Choi:1998ia}.

The total and relative fluctuations are given by the unitary transformation,
    \begin{equation}
    \begin{split}
        \updelta\hat{\psi}_\theta&=\sqrt{\frac{1+z}{2}}\updelta\hat{\psi}_\down+\sqrt{\frac{1-z}{2}}\updelta\hat{\psi}_\up,\\
        \updelta\hat{\psi}_\phi&=\sqrt{\frac{1-z}{2}}\updelta\hat{\psi}_\down-\sqrt{\frac{1+z}{2}}\updelta\hat{\psi}_\up.
    \end{split}
    \end{equation}
For each sector we carry out a Bogoliubov transformation,
    \begin{equation}
    \label{eq:mode-expansion}
        \updelta\hat{\psi}(\vb*x)=\sum_i\qty[u_i^{}(\vb*x)\hat{a}_i^{}-v_i^{}(\vb*x)\hat{a}_i^\dagger],
    \end{equation}
    where $i$ labels normal modes of the system, and the mode functions $u_i$, $v_i$ are chosen such that they diagonalize the Hamiltonian~\eqref{eq:hamiltonian-time-averaged},
    \begin{equation}
    \label{eq:quadratic-hamiltonian}
        \hat{H}=\int\dd{\vb*x}\hat{\mathcal{H}}\simeq E_0+\sum_i\qty[\hbar\omega_{\theta,i}^{}\hat{a}_{\theta,i}^\dagger\hat{a}_{\theta,i}^{}+\hbar\omega_{\phi,i}^{}\hat{a}_{\phi,i}^\dagger\hat{a}_{\phi,i}^{}],
    \end{equation}
    where we have expanded up to quadratic order in the fluctuations $\updelta\hat{\psi}$.
(The constant background energy $E_0$ can be ignored, as it does not influence the mode functions.)
Neglecting higher-order terms here corresponds to approximating the modes as non-interacting, leading to Gaussian fluctuation statistics.
The operators $\hat{a}^\dagger$, $\hat{a}$ are then the standard creation and annihilation operators for each mode, and are treated as i.i.d. classical random variables drawn from a complex, uniform-phase Gaussian distribution with zero mean and variance $1/2$, following the usual truncated Wigner prescription.

\begin{figure}[t!]
    \centering
    \includegraphics{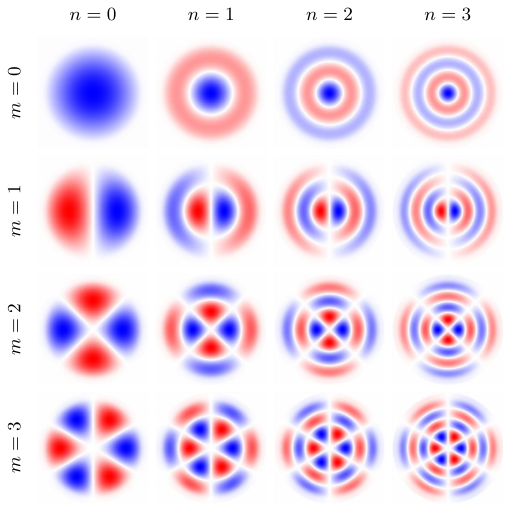}
    \caption{\label{fig:bogoliubov}
    The first few eigenmodes for relative phase fluctuations in the trench potential.
    We show $u_{mn}(\vb*x)$ with arbitrary normalization, for illustrative purposes; the corresponding $v_{mn}(\vb*x)$ are qualitatively very similar.
    Blue and red correspond to positive and negative values, respectively.}
\end{figure}

Combining Eqs.~\eqref{eq:mode-expansion} and~\eqref{eq:quadratic-hamiltonian}, we find the Bogoliubov equations that determine the mode functions for each fluctuation sector, which are of the form
    \begin{equation}
    \label{eq:coupled-system}
    \begin{split}
        \hbar\omega_i u_i&=\mathcal{A}u_i-\mathcal{B}v_i,\\
        -\hbar\omega_i v_i&=\mathcal{A}v_i-\mathcal{B}u_i,
    \end{split}
    \end{equation}
    where $\mathcal{A}$, $\mathcal{B}$ are linear differential operators.
The solutions to this system are normalized according to
    \begin{equation}
    \label{eq:normalization}
        \int\dd{\vb*x}(u_i u_j^*-v_i v_j^*)=\delta_{ij}
    \end{equation}
    to ensure the ladder operators obey the usual commutation relations.

The coupled system~\eqref{eq:coupled-system} is somewhat awkward to solve directly.
Instead, it is convenient to take the odd and even combinations,
    \begin{equation}
    \begin{split}
        \hbar\omega_i u_{+,i}&=\mathcal{L}_+u_{-,i},\\
        \hbar\omega_i u_{-,i}&=\mathcal{L}_-u_{+,i},
    \end{split}
    \end{equation}
    where we define $u_{\pm,i}=u_i\pm v_i$ and $\mathcal{L}_\pm=\mathcal{A}\pm\mathcal{B}$.
Chaining these equations together yields
    \begin{equation}
    \begin{split}
        {(\hbar\omega_i)}^2u_{+,i}&=\mathcal{L}_+\mathcal{L}_-u_{+,i},\\
        {(\hbar\omega_i)}^2u_{-,i}&=\mathcal{L}_-\mathcal{L}_+u_{-,i},
    \end{split}
    \end{equation}
    each of which is a self-contained eigenvalue problem that is amenable to solution via standard numerical methods.
Our procedure is therefore: (1) find the eigenvalues ${(\hbar\omega_i)}^2$ and eigenfunctions $u_{+,i}$ of the operator $\mathcal{L}_+\mathcal{L}_-$; (2) for each $u_{+,i}$, apply the operator $\mathcal{L}_-$ and divide by $\hbar\omega_i$ to find the corresponding $u_{-,i}$; (3) take odd and even combinations and enforce the normalization~\eqref{eq:normalization} to find the mode functions $u_i$, $v_i$.

\begin{figure}[t!]
    \centering
    \includegraphics{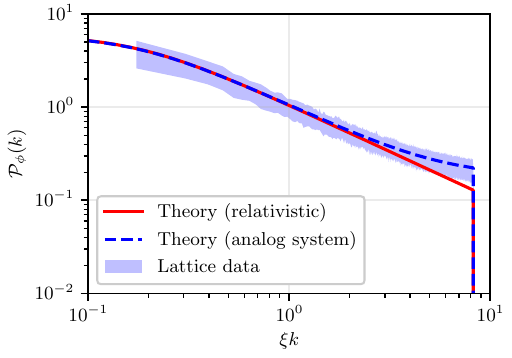}
    \caption{\label{fig:spectrum}
    Initial power spectrum of the relative phase field in our simulations (shaded region, which shows $\pm1\sigma$ around the estimated spectrum), as estimated from the bulk region of our high-density trench ensemble at time zero.
    We find excellent agreement with the expected spectrum for a periodic analog system (dashed blue curve), including the UV cutoff at $k\approx8.25/\xi$ and the slight excess UV power compared to the corresponding relativistic theory (solid red curve).}
\end{figure}

Since we are working on the lattice, step (1) above involves approximating the operators $\mathcal{L}_\pm$ as matrices acting on vectors that specify $u_\pm$ at each lattice site.
Naively, for a 2D lattice with $N^2$ sites, this means diagonalizing a matrix of size $N^2\times N^2$ and storing the resulting $N^2$ eigenvectors.
This is infeasible for $N=1024$.
Instead, we exploit the circular symmetry of the system and work in polar coordinates $(r,\theta)$.
The mode functions can then be written as
    \begin{equation}
    \begin{split}
        u_i(\vb*x)=u_{mn}(\vb*x)=U_{mn}(r)\rme^{\rmi m\theta},\\
        v_i(\vb*x)=v_{mn}(\vb*x)=V_{mn}(r)\rme^{\rmi m\theta},
    \end{split}
    \end{equation}
    \newline
    where $m=0,\pm1,\pm2,\ldots$ (not to be confused with the atomic mass) labels modes of different angular momenta, and $n=0,1,2,\ldots$ (not to be confused with the number density) labels energy levels for each $m$.
The problem then reduces to solving a radial eigenvalue equation for each $m$, each described in terms of an $N\times N$ matrix.
We generate these matrices using a pseudospectral representation for the differential operators $\mathcal{L}_\pm$, and use the same pseudospectral scheme to interpolate the resulting radial mode functions on the Cartesian lattice used in the simulations.
This procedure is carried out separately for each of the two trapping potentials described in Sec.~\ref{sec:sims}; Fig.~\ref{fig:bogoliubov} shows the first few modes in the trench case.

We perform two tests to confirm that the resulting Bogoliubov modes accurately describe the vacuum fluctuations of the system.
First, we carry out a truncated Wigner simulation with very small initial fluctuation amplitudes, corresponding to $N\xi^2/A=10^8$.
Interactions between modes should be negligible in this regime, so that each mode simply oscillates at its natural frequency,
    \begin{equation}
    \label{eq:linear-evolution}
        \hat{a}_{mn}(t)\simeq\hat{a}_{mn}(0)\,\rme^{-\rmi\omega_{mn}t}.
    \end{equation}
We extract the mode amplitudes from the simulation and find that they each obey~\eqref{eq:linear-evolution} with relative accuracy of $10^{-3}$ or better over many oscillation periods.
This confirms that the modes diagonalize the Hamiltonian to high accuracy in the linear regime, and that the energy eigenvalues $\hbar\omega_{mn}$ are accurate.

Our second test is to compute the initial power spectrum of the relative phase field,
    \begin{equation}
        \mathcal{P}_\phi(k)=\int\frac{\dd{\vb*x}}{V}\,\rme^{\rmi\vb*k\vdot(\vb*x'-\vb*x)}\ev{\phi(\vb*x)\phi(\vb*x')}.
    \end{equation}
We do this by averaging over the 512 simulations in our high-density trench ensemble, focusing on a square subregion of side length $\approx80\,\xi$ that is contained entirely within the bulk, and using a Slepian window to suppress spectral leakage.
As shown in Fig.~\ref{fig:spectrum}, we find excellent agreement with the expected spectrum.
This demonstrates that our numerical framework accurately reproduces the fluctuation statistics of the corresponding periodic system in the bulk region of the trap.

%%%%%%%%%%%%%%%%%%%%%%%%%%%%%%%%%%%%%%%%%%%%%%%%%%%%%%%%%%%%%%%%%%%%%%%%%%%%%%%
\bibliography{bubbles-in-a-box}

\begin{thebibliography}{83}%
\makeatletter
\providecommand \@ifxundefined [1]{%
 \@ifx{#1\undefined}
}%
\providecommand \@ifnum [1]{%
 \ifnum #1\expandafter \@firstoftwo
 \else \expandafter \@secondoftwo
 \fi
}%
\providecommand \@ifx [1]{%
 \ifx #1\expandafter \@firstoftwo
 \else \expandafter \@secondoftwo
 \fi
}%
\providecommand \natexlab [1]{#1}%
\providecommand \enquote  [1]{``#1''}%
\providecommand \bibnamefont  [1]{#1}%
\providecommand \bibfnamefont [1]{#1}%
\providecommand \citenamefont [1]{#1}%
\providecommand \href@noop [0]{\@secondoftwo}%
\providecommand \href [0]{\begingroup \@sanitize@url \@href}%
\providecommand \@href[1]{\@@startlink{#1}\@@href}%
\providecommand \@@href[1]{\endgroup#1\@@endlink}%
\providecommand \@sanitize@url [0]{\catcode `\\12\catcode `\$12\catcode
  `\&12\catcode `\#12\catcode `\^12\catcode `\_12\catcode `\%12\relax}%
\providecommand \@@startlink[1]{}%
\providecommand \@@endlink[0]{}%
\providecommand \url  [0]{\begingroup\@sanitize@url \@url }%
\providecommand \@url [1]{\endgroup\@href {#1}{\urlprefix }}%
\providecommand \urlprefix  [0]{URL }%
\providecommand \Eprint [0]{\href }%
\providecommand \doibase [0]{https://doi.org/}%
\providecommand \selectlanguage [0]{\@gobble}%
\providecommand \bibinfo  [0]{\@secondoftwo}%
\providecommand \bibfield  [0]{\@secondoftwo}%
\providecommand \translation [1]{[#1]}%
\providecommand \BibitemOpen [0]{}%
\providecommand \bibitemStop [0]{}%
\providecommand \bibitemNoStop [0]{.\EOS\space}%
\providecommand \EOS [0]{\spacefactor3000\relax}%
\providecommand \BibitemShut  [1]{\csname bibitem#1\endcsname}%
\let\auto@bib@innerbib\@empty
%</preamble>
\bibitem [{\citenamefont {Opanchuk}\ \emph {et~al.}(2013)\citenamefont
  {Opanchuk}, \citenamefont {Polkinghorne}, \citenamefont {Fialko},
  \citenamefont {Brand},\ and\ \citenamefont {Drummond}}]{Opanchuk:2013lgn}%
  \BibitemOpen
  \bibfield  {author} {\bibinfo {author} {\bibfnamefont {B.}~\bibnamefont
  {Opanchuk}}, \bibinfo {author} {\bibfnamefont {R.}~\bibnamefont
  {Polkinghorne}}, \bibinfo {author} {\bibfnamefont {O.}~\bibnamefont
  {Fialko}}, \bibinfo {author} {\bibfnamefont {J.}~\bibnamefont {Brand}},\ and\
  \bibinfo {author} {\bibfnamefont {P.~D.}\ \bibnamefont {Drummond}},\
  }\bibfield  {title} {\bibinfo {title} {{Quantum simulations of the early
  universe}},\ }\href {https://doi.org/10.1002/andp.201300113} {\bibfield
  {journal} {\bibinfo  {journal} {Annalen Phys.}\ }\textbf {\bibinfo {volume}
  {525}},\ \bibinfo {pages} {866} (\bibinfo {year} {2013})},\ \Eprint
  {https://arxiv.org/abs/1305.5314} {arXiv:1305.5314 [cond-mat.quant-gas]}
  \BibitemShut {NoStop}%
\bibitem [{\citenamefont {Fialko}\ \emph {et~al.}(2015)\citenamefont {Fialko},
  \citenamefont {Opanchuk}, \citenamefont {Sidorov}, \citenamefont {Drummond},\
  and\ \citenamefont {Brand}}]{Fialko:2014xba}%
  \BibitemOpen
  \bibfield  {author} {\bibinfo {author} {\bibfnamefont {O.}~\bibnamefont
  {Fialko}}, \bibinfo {author} {\bibfnamefont {B.}~\bibnamefont {Opanchuk}},
  \bibinfo {author} {\bibfnamefont {A.~I.}\ \bibnamefont {Sidorov}}, \bibinfo
  {author} {\bibfnamefont {P.~D.}\ \bibnamefont {Drummond}},\ and\ \bibinfo
  {author} {\bibfnamefont {J.}~\bibnamefont {Brand}},\ }\bibfield  {title}
  {\bibinfo {title} {{Fate of the false vacuum: towards realization with
  ultra-cold atoms}},\ }\href {https://doi.org/10.1209/0295-5075/110/56001}
  {\bibfield  {journal} {\bibinfo  {journal} {EPL}\ }\textbf {\bibinfo {volume}
  {110}},\ \bibinfo {pages} {56001} (\bibinfo {year} {2015})},\ \Eprint
  {https://arxiv.org/abs/1408.1163} {arXiv:1408.1163 [cond-mat.quant-gas]}
  \BibitemShut {NoStop}%
\bibitem [{\citenamefont {Fialko}\ \emph {et~al.}(2017)\citenamefont {Fialko},
  \citenamefont {Opanchuk}, \citenamefont {Sidorov}, \citenamefont {Drummond},\
  and\ \citenamefont {Brand}}]{Fialko:2016ggg}%
  \BibitemOpen
  \bibfield  {author} {\bibinfo {author} {\bibfnamefont {O.}~\bibnamefont
  {Fialko}}, \bibinfo {author} {\bibfnamefont {B.}~\bibnamefont {Opanchuk}},
  \bibinfo {author} {\bibfnamefont {A.~I.}\ \bibnamefont {Sidorov}}, \bibinfo
  {author} {\bibfnamefont {P.~D.}\ \bibnamefont {Drummond}},\ and\ \bibinfo
  {author} {\bibfnamefont {J.}~\bibnamefont {Brand}},\ }\bibfield  {title}
  {\bibinfo {title} {{The universe on a table top: engineering quantum decay of
  a relativistic scalar field from a metastable vacuum}},\ }\href
  {https://doi.org/10.1088/1361-6455/50/2/024003} {\bibfield  {journal}
  {\bibinfo  {journal} {J. Phys. B}\ }\textbf {\bibinfo {volume} {50}},\
  \bibinfo {pages} {024003} (\bibinfo {year} {2017})},\ \Eprint
  {https://arxiv.org/abs/1607.01460} {arXiv:1607.01460 [cond-mat.quant-gas]}
  \BibitemShut {NoStop}%
\bibitem [{\citenamefont {Braden}\ \emph {et~al.}(2018)\citenamefont {Braden},
  \citenamefont {Johnson}, \citenamefont {Peiris},\ and\ \citenamefont
  {Weinfurtner}}]{Braden:2017add}%
  \BibitemOpen
  \bibfield  {author} {\bibinfo {author} {\bibfnamefont {J.}~\bibnamefont
  {Braden}}, \bibinfo {author} {\bibfnamefont {M.~C.}\ \bibnamefont {Johnson}},
  \bibinfo {author} {\bibfnamefont {H.~V.}\ \bibnamefont {Peiris}},\ and\
  \bibinfo {author} {\bibfnamefont {S.}~\bibnamefont {Weinfurtner}},\
  }\bibfield  {title} {\bibinfo {title} {{Towards the cold atom analog false
  vacuum}},\ }\href {https://doi.org/10.1007/JHEP07(2018)014} {\bibfield
  {journal} {\bibinfo  {journal} {JHEP}\ }\textbf {\bibinfo {volume}
  {07}}\bibfield  {number} {\bibinfo  {number} { (2018)},\ \bibinfo {pages}
  {014}},\ }\Eprint {https://arxiv.org/abs/1712.02356} {arXiv:1712.02356
  [hep-th]} \BibitemShut {NoStop}%
\bibitem [{\citenamefont {Billam}\ \emph {et~al.}(2019)\citenamefont {Billam},
  \citenamefont {Gregory}, \citenamefont {Michel},\ and\ \citenamefont
  {Moss}}]{Billam:2018pvp}%
  \BibitemOpen
  \bibfield  {author} {\bibinfo {author} {\bibfnamefont {T.~P.}\ \bibnamefont
  {Billam}}, \bibinfo {author} {\bibfnamefont {R.}~\bibnamefont {Gregory}},
  \bibinfo {author} {\bibfnamefont {F.}~\bibnamefont {Michel}},\ and\ \bibinfo
  {author} {\bibfnamefont {I.~G.}\ \bibnamefont {Moss}},\ }\bibfield  {title}
  {\bibinfo {title} {{Simulating seeded vacuum decay in a cold atom system}},\
  }\href {https://doi.org/10.1103/PhysRevD.100.065016} {\bibfield  {journal}
  {\bibinfo  {journal} {Phys. Rev. D}\ }\textbf {\bibinfo {volume} {100}},\
  \bibinfo {pages} {065016} (\bibinfo {year} {2019})},\ \Eprint
  {https://arxiv.org/abs/1811.09169} {arXiv:1811.09169 [hep-th]} \BibitemShut
  {NoStop}%
\bibitem [{\citenamefont {Braden}\ \emph
  {et~al.}(2019{\natexlab{a}})\citenamefont {Braden}, \citenamefont {Johnson},
  \citenamefont {Peiris}, \citenamefont {Pontzen},\ and\ \citenamefont
  {Weinfurtner}}]{Braden:2019vsw}%
  \BibitemOpen
  \bibfield  {author} {\bibinfo {author} {\bibfnamefont {J.}~\bibnamefont
  {Braden}}, \bibinfo {author} {\bibfnamefont {M.~C.}\ \bibnamefont {Johnson}},
  \bibinfo {author} {\bibfnamefont {H.~V.}\ \bibnamefont {Peiris}}, \bibinfo
  {author} {\bibfnamefont {A.}~\bibnamefont {Pontzen}},\ and\ \bibinfo {author}
  {\bibfnamefont {S.}~\bibnamefont {Weinfurtner}},\ }\bibfield  {title}
  {\bibinfo {title} {{Nonlinear Dynamics of the Cold Atom Analog False
  Vacuum}},\ }\href {https://doi.org/10.1007/JHEP10(2019)174} {\bibfield
  {journal} {\bibinfo  {journal} {JHEP}\ }\textbf {\bibinfo {volume}
  {10}}\bibfield  {number} {\bibinfo  {number} { (2019)},\ \bibinfo {pages}
  {174}},\ }\Eprint {https://arxiv.org/abs/1904.07873} {arXiv:1904.07873
  [hep-th]} \BibitemShut {NoStop}%
\bibitem [{\citenamefont {Billam}\ \emph {et~al.}(2020)\citenamefont {Billam},
  \citenamefont {Brown},\ and\ \citenamefont {Moss}}]{Billam:2020xna}%
  \BibitemOpen
  \bibfield  {author} {\bibinfo {author} {\bibfnamefont {T.~P.}\ \bibnamefont
  {Billam}}, \bibinfo {author} {\bibfnamefont {K.}~\bibnamefont {Brown}},\ and\
  \bibinfo {author} {\bibfnamefont {I.~G.}\ \bibnamefont {Moss}},\ }\bibfield
  {title} {\bibinfo {title} {{Simulating cosmological supercooling with a cold
  atom system}},\ }\href {https://doi.org/10.1103/PhysRevA.102.043324}
  {\bibfield  {journal} {\bibinfo  {journal} {Phys. Rev. A}\ }\textbf {\bibinfo
  {volume} {102}},\ \bibinfo {pages} {043324} (\bibinfo {year} {2020})},\
  \Eprint {https://arxiv.org/abs/2006.09820} {arXiv:2006.09820
  [cond-mat.quant-gas]} \BibitemShut {NoStop}%
\bibitem [{\citenamefont {Ng}\ \emph {et~al.}(2021)\citenamefont {Ng},
  \citenamefont {Opanchuk}, \citenamefont {Thenabadu}, \citenamefont {Reid},\
  and\ \citenamefont {Drummond}}]{Ng:2020pxk}%
  \BibitemOpen
  \bibfield  {author} {\bibinfo {author} {\bibfnamefont {K.~L.}\ \bibnamefont
  {Ng}}, \bibinfo {author} {\bibfnamefont {B.}~\bibnamefont {Opanchuk}},
  \bibinfo {author} {\bibfnamefont {M.}~\bibnamefont {Thenabadu}}, \bibinfo
  {author} {\bibfnamefont {M.}~\bibnamefont {Reid}},\ and\ \bibinfo {author}
  {\bibfnamefont {P.~D.}\ \bibnamefont {Drummond}},\ }\bibfield  {title}
  {\bibinfo {title} {{The fate of the false vacuum: Finite temperature, entropy
  and topological phase in quantum simulations of the early universe}},\ }\href
  {https://doi.org/10.1103/PRXQuantum.2.010350} {\bibfield  {journal} {\bibinfo
   {journal} {PRX Quantum}\ }\textbf {\bibinfo {volume} {2}},\ \bibinfo {pages}
  {010350} (\bibinfo {year} {2021})},\ \Eprint
  {https://arxiv.org/abs/2010.08665} {arXiv:2010.08665 [quant-ph]} \BibitemShut
  {NoStop}%
\bibitem [{\citenamefont {Billam}\ \emph {et~al.}(2021)\citenamefont {Billam},
  \citenamefont {Brown}, \citenamefont {Groszek},\ and\ \citenamefont
  {Moss}}]{Billam:2021qwt}%
  \BibitemOpen
  \bibfield  {author} {\bibinfo {author} {\bibfnamefont {T.~P.}\ \bibnamefont
  {Billam}}, \bibinfo {author} {\bibfnamefont {K.}~\bibnamefont {Brown}},
  \bibinfo {author} {\bibfnamefont {A.~J.}\ \bibnamefont {Groszek}},\ and\
  \bibinfo {author} {\bibfnamefont {I.~G.}\ \bibnamefont {Moss}},\ }\bibfield
  {title} {\bibinfo {title} {{Simulating cosmological supercooling with a cold
  atom system. II. Thermal damping and parametric instability}},\ }\href
  {https://doi.org/10.1103/PhysRevA.104.053309} {\bibfield  {journal} {\bibinfo
   {journal} {Phys. Rev. A}\ }\textbf {\bibinfo {volume} {104}},\ \bibinfo
  {pages} {053309} (\bibinfo {year} {2021})},\ \Eprint
  {https://arxiv.org/abs/2104.07428} {arXiv:2104.07428 [cond-mat.quant-gas]}
  \BibitemShut {NoStop}%
\bibitem [{\citenamefont {Billam}\ \emph {et~al.}(2022)\citenamefont {Billam},
  \citenamefont {Brown},\ and\ \citenamefont {Moss}}]{Billam:2021nbc}%
  \BibitemOpen
  \bibfield  {author} {\bibinfo {author} {\bibfnamefont {T.~P.}\ \bibnamefont
  {Billam}}, \bibinfo {author} {\bibfnamefont {K.}~\bibnamefont {Brown}},\ and\
  \bibinfo {author} {\bibfnamefont {I.~G.}\ \bibnamefont {Moss}},\ }\bibfield
  {title} {\bibinfo {title} {{False-vacuum decay in an ultracold spin-1 Bose
  gas}},\ }\href {https://doi.org/10.1103/PhysRevA.105.L041301} {\bibfield
  {journal} {\bibinfo  {journal} {Phys. Rev. A}\ }\textbf {\bibinfo {volume}
  {105}},\ \bibinfo {pages} {L041301} (\bibinfo {year} {2022})},\ \Eprint
  {https://arxiv.org/abs/2108.05740} {arXiv:2108.05740 [cond-mat.quant-gas]}
  \BibitemShut {NoStop}%
\bibitem [{\citenamefont {Billam}\ \emph {et~al.}(2023)\citenamefont {Billam},
  \citenamefont {Brown},\ and\ \citenamefont {Moss}}]{Billam:2022ykl}%
  \BibitemOpen
  \bibfield  {author} {\bibinfo {author} {\bibfnamefont {T.~P.}\ \bibnamefont
  {Billam}}, \bibinfo {author} {\bibfnamefont {K.}~\bibnamefont {Brown}},\ and\
  \bibinfo {author} {\bibfnamefont {I.~G.}\ \bibnamefont {Moss}},\ }\bibfield
  {title} {\bibinfo {title} {{Bubble nucleation in a cold spin 1 gas}},\ }\href
  {https://doi.org/10.1088/1367-2630/accca2} {\bibfield  {journal} {\bibinfo
  {journal} {New J. Phys.}\ }\textbf {\bibinfo {volume} {25}},\ \bibinfo
  {pages} {043028} (\bibinfo {year} {2023})},\ \Eprint
  {https://arxiv.org/abs/2212.03621} {arXiv:2212.03621 [cond-mat.quant-gas]}
  \BibitemShut {NoStop}%
\bibitem [{\citenamefont {Jenkins}\ \emph
  {et~al.}(2024{\natexlab{a}})\citenamefont {Jenkins}, \citenamefont {Braden},
  \citenamefont {Peiris}, \citenamefont {Pontzen}, \citenamefont {Johnson},\
  and\ \citenamefont {Weinfurtner}}]{Jenkins:2023eez}%
  \BibitemOpen
  \bibfield  {author} {\bibinfo {author} {\bibfnamefont {A.~C.}\ \bibnamefont
  {Jenkins}}, \bibinfo {author} {\bibfnamefont {J.}~\bibnamefont {Braden}},
  \bibinfo {author} {\bibfnamefont {H.~V.}\ \bibnamefont {Peiris}}, \bibinfo
  {author} {\bibfnamefont {A.}~\bibnamefont {Pontzen}}, \bibinfo {author}
  {\bibfnamefont {M.~C.}\ \bibnamefont {Johnson}},\ and\ \bibinfo {author}
  {\bibfnamefont {S.}~\bibnamefont {Weinfurtner}},\ }\bibfield  {title}
  {\bibinfo {title} {{Analog vacuum decay from vacuum initial conditions}},\
  }\href {https://doi.org/10.1103/PhysRevD.109.023506} {\bibfield  {journal}
  {\bibinfo  {journal} {Phys. Rev. D}\ }\textbf {\bibinfo {volume} {109}},\
  \bibinfo {pages} {023506} (\bibinfo {year} {2024}{\natexlab{a}})},\ \Eprint
  {https://arxiv.org/abs/2307.02549} {arXiv:2307.02549 [cond-mat.quant-gas]}
  \BibitemShut {NoStop}%
\bibitem [{\citenamefont {Jenkins}\ \emph
  {et~al.}(2024{\natexlab{b}})\citenamefont {Jenkins}, \citenamefont {Moss},
  \citenamefont {Billam}, \citenamefont {Hadzibabic}, \citenamefont {Peiris},\
  and\ \citenamefont {Pontzen}}]{Jenkins:2023npg}%
  \BibitemOpen
  \bibfield  {author} {\bibinfo {author} {\bibfnamefont {A.~C.}\ \bibnamefont
  {Jenkins}}, \bibinfo {author} {\bibfnamefont {I.~G.}\ \bibnamefont {Moss}},
  \bibinfo {author} {\bibfnamefont {T.~P.}\ \bibnamefont {Billam}}, \bibinfo
  {author} {\bibfnamefont {Z.}~\bibnamefont {Hadzibabic}}, \bibinfo {author}
  {\bibfnamefont {H.~V.}\ \bibnamefont {Peiris}},\ and\ \bibinfo {author}
  {\bibfnamefont {A.}~\bibnamefont {Pontzen}},\ }\bibfield  {title} {\bibinfo
  {title} {{Generalized cold-atom simulators for vacuum decay}},\ }\href
  {https://doi.org/10.1103/PhysRevA.110.L031301} {\bibfield  {journal}
  {\bibinfo  {journal} {Phys. Rev. A}\ }\textbf {\bibinfo {volume} {110}},\
  \bibinfo {pages} {L031301} (\bibinfo {year} {2024}{\natexlab{b}})},\ \Eprint
  {https://arxiv.org/abs/2311.02156} {arXiv:2311.02156 [cond-mat.quant-gas]}
  \BibitemShut {NoStop}%
\bibitem [{\citenamefont {Zenesini}\ \emph {et~al.}(2024)\citenamefont
  {Zenesini}, \citenamefont {Berti}, \citenamefont {Cominotti}, \citenamefont
  {Rogora}, \citenamefont {Moss}, \citenamefont {Billam}, \citenamefont
  {Carusotto}, \citenamefont {Lamporesi}, \citenamefont {Recati},\ and\
  \citenamefont {Ferrari}}]{Zenesini:2023afv}%
  \BibitemOpen
  \bibfield  {author} {\bibinfo {author} {\bibfnamefont {A.}~\bibnamefont
  {Zenesini}}, \bibinfo {author} {\bibfnamefont {A.}~\bibnamefont {Berti}},
  \bibinfo {author} {\bibfnamefont {R.}~\bibnamefont {Cominotti}}, \bibinfo
  {author} {\bibfnamefont {C.}~\bibnamefont {Rogora}}, \bibinfo {author}
  {\bibfnamefont {I.~G.}\ \bibnamefont {Moss}}, \bibinfo {author}
  {\bibfnamefont {T.~P.}\ \bibnamefont {Billam}}, \bibinfo {author}
  {\bibfnamefont {I.}~\bibnamefont {Carusotto}}, \bibinfo {author}
  {\bibfnamefont {G.}~\bibnamefont {Lamporesi}}, \bibinfo {author}
  {\bibfnamefont {A.}~\bibnamefont {Recati}},\ and\ \bibinfo {author}
  {\bibfnamefont {G.}~\bibnamefont {Ferrari}},\ }\bibfield  {title} {\bibinfo
  {title} {{False vacuum decay via bubble formation in ferromagnetic
  superfluids}},\ }\href {https://doi.org/10.1038/s41567-023-02345-4}
  {\bibfield  {journal} {\bibinfo  {journal} {Nature Phys.}\ }\textbf {\bibinfo
  {volume} {20}},\ \bibinfo {pages} {558} (\bibinfo {year} {2024})},\ \Eprint
  {https://arxiv.org/abs/2305.05225} {arXiv:2305.05225 [hep-ph]} \BibitemShut
  {NoStop}%
\bibitem [{\citenamefont {Cominotti}\ \emph {et~al.}(2025)\citenamefont
  {Cominotti}, \citenamefont {Baroni}, \citenamefont {Rogora}, \citenamefont
  {Andreoni}, \citenamefont {Guarda}, \citenamefont {Lamporesi}, \citenamefont
  {Ferrari},\ and\ \citenamefont {Zenesini}}]{Cominotti:2025qia}%
  \BibitemOpen
  \bibfield  {author} {\bibinfo {author} {\bibfnamefont {R.}~\bibnamefont
  {Cominotti}}, \bibinfo {author} {\bibfnamefont {C.}~\bibnamefont {Baroni}},
  \bibinfo {author} {\bibfnamefont {C.}~\bibnamefont {Rogora}}, \bibinfo
  {author} {\bibfnamefont {D.}~\bibnamefont {Andreoni}}, \bibinfo {author}
  {\bibfnamefont {G.}~\bibnamefont {Guarda}}, \bibinfo {author} {\bibfnamefont
  {G.}~\bibnamefont {Lamporesi}}, \bibinfo {author} {\bibfnamefont
  {G.}~\bibnamefont {Ferrari}},\ and\ \bibinfo {author} {\bibfnamefont
  {A.}~\bibnamefont {Zenesini}},\ }\bibfield  {title} {\bibinfo {title}
  {{Observation of Temperature Effects in False Vacuum Decay}},\ }\href@noop {}
  {\bibfield  {journal} {\bibinfo  {journal} {{}}\ } (\bibinfo {year}
  {2025})},\ \Eprint {https://arxiv.org/abs/2504.03528} {arXiv:2504.03528
  [cond-mat.quant-gas]} \BibitemShut {NoStop}%
\bibitem [{\citenamefont {Fischer}\ and\ \citenamefont
  {Sch{\"u}tzhold}(2004)}]{Fischer:2004bf}%
  \BibitemOpen
  \bibfield  {author} {\bibinfo {author} {\bibfnamefont {U.~R.}\ \bibnamefont
  {Fischer}}\ and\ \bibinfo {author} {\bibfnamefont {R.}~\bibnamefont
  {Sch{\"u}tzhold}},\ }\bibfield  {title} {\bibinfo {title} {{Quantum
  simulation of cosmic inflation in two-component Bose-Einstein condensates}},\
  }\href {https://doi.org/10.1103/PhysRevA.70.063615} {\bibfield  {journal}
  {\bibinfo  {journal} {Phys. Rev. A}\ }\textbf {\bibinfo {volume} {70}},\
  \bibinfo {pages} {063615} (\bibinfo {year} {2004})},\ \Eprint
  {https://arxiv.org/abs/cond-mat/0406470} {arXiv:cond-mat/0406470}
  \BibitemShut {NoStop}%
\bibitem [{\citenamefont {Visser}\ and\ \citenamefont
  {Weinfurtner}(2004)}]{Visser:2004qp}%
  \BibitemOpen
  \bibfield  {author} {\bibinfo {author} {\bibfnamefont {M.}~\bibnamefont
  {Visser}}\ and\ \bibinfo {author} {\bibfnamefont {S.}~\bibnamefont
  {Weinfurtner}},\ }\bibfield  {title} {\bibinfo {title} {{Massive phonon modes
  from a BEC-based analog model}},\ }\href@noop {} {\bibfield  {journal}
  {\bibinfo  {journal} {{}}\ } (\bibinfo {year} {2004})},\ \Eprint
  {https://arxiv.org/abs/cond-mat/0409639} {arXiv:cond-mat/0409639}
  \BibitemShut {NoStop}%
\bibitem [{\citenamefont {Visser}\ and\ \citenamefont
  {Weinfurtner}(2005)}]{Visser:2005ss}%
  \BibitemOpen
  \bibfield  {author} {\bibinfo {author} {\bibfnamefont {M.}~\bibnamefont
  {Visser}}\ and\ \bibinfo {author} {\bibfnamefont {S.}~\bibnamefont
  {Weinfurtner}},\ }\bibfield  {title} {\bibinfo {title} {{Massive Klein-Gordon
  equation from a BEC-based analogue spacetime}},\ }\href
  {https://doi.org/10.1103/PhysRevD.72.044020} {\bibfield  {journal} {\bibinfo
  {journal} {Phys. Rev. D}\ }\textbf {\bibinfo {volume} {72}},\ \bibinfo
  {pages} {044020} (\bibinfo {year} {2005})},\ \Eprint
  {https://arxiv.org/abs/gr-qc/0506029} {arXiv:gr-qc/0506029} \BibitemShut
  {NoStop}%
\bibitem [{\citenamefont {Weinfurtner}\ \emph {et~al.}(2007)\citenamefont
  {Weinfurtner}, \citenamefont {Liberati},\ and\ \citenamefont
  {Visser}}]{Weinfurtner:2006wt}%
  \BibitemOpen
  \bibfield  {author} {\bibinfo {author} {\bibfnamefont {S.}~\bibnamefont
  {Weinfurtner}}, \bibinfo {author} {\bibfnamefont {S.}~\bibnamefont
  {Liberati}},\ and\ \bibinfo {author} {\bibfnamefont {M.}~\bibnamefont
  {Visser}},\ }\bibfield  {title} {\bibinfo {title} {{Analogue spacetime based
  on 2-component Bose-Einstein condensates}},\ }\href
  {https://doi.org/10.1007/3-540-70859-6_6} {\bibfield  {journal} {\bibinfo
  {journal} {Lect. Notes Phys.}\ }\textbf {\bibinfo {volume} {718}},\ \bibinfo
  {pages} {115} (\bibinfo {year} {2007})},\ \Eprint
  {https://arxiv.org/abs/gr-qc/0605121} {arXiv:gr-qc/0605121} \BibitemShut
  {NoStop}%
\bibitem [{\citenamefont {Neuenhahn}\ and\ \citenamefont
  {Marquardt}(2015)}]{Neuenhahn:2012dz}%
  \BibitemOpen
  \bibfield  {author} {\bibinfo {author} {\bibfnamefont {C.}~\bibnamefont
  {Neuenhahn}}\ and\ \bibinfo {author} {\bibfnamefont {F.}~\bibnamefont
  {Marquardt}},\ }\bibfield  {title} {\bibinfo {title} {{Quantum simulation of
  expanding space-time with tunnel-coupled condensates}},\ }\href
  {https://doi.org/10.1088/1367-2630/17/12/125007} {\bibfield  {journal}
  {\bibinfo  {journal} {New J. Phys.}\ }\textbf {\bibinfo {volume} {17}},\
  \bibinfo {pages} {125007} (\bibinfo {year} {2015})},\ \Eprint
  {https://arxiv.org/abs/1208.2255} {arXiv:1208.2255 [cond-mat.quant-gas]}
  \BibitemShut {NoStop}%
\bibitem [{\citenamefont {Su}\ \emph {et~al.}(2015)\citenamefont {Su},
  \citenamefont {Gou}, \citenamefont {Liu}, \citenamefont {Bradley},
  \citenamefont {Fialko},\ and\ \citenamefont {Brand}}]{Su:2014osc}%
  \BibitemOpen
  \bibfield  {author} {\bibinfo {author} {\bibfnamefont {S.-W.}\ \bibnamefont
  {Su}}, \bibinfo {author} {\bibfnamefont {S.-C.}\ \bibnamefont {Gou}},
  \bibinfo {author} {\bibfnamefont {I.-K.}\ \bibnamefont {Liu}}, \bibinfo
  {author} {\bibfnamefont {A.~S.}\ \bibnamefont {Bradley}}, \bibinfo {author}
  {\bibfnamefont {O.}~\bibnamefont {Fialko}},\ and\ \bibinfo {author}
  {\bibfnamefont {J.}~\bibnamefont {Brand}},\ }\bibfield  {title} {\bibinfo
  {title} {{Oscillons in coupled Bose-Einstein condensates}},\ }\href
  {https://doi.org/10.1103/PhysRevA.91.023631} {\bibfield  {journal} {\bibinfo
  {journal} {Phys. Rev. A}\ }\textbf {\bibinfo {volume} {91}},\ \bibinfo
  {pages} {023631} (\bibinfo {year} {2015})},\ \Eprint
  {https://arxiv.org/abs/1412.5858} {arXiv:1412.5858 [cond-mat.quant-gas]}
  \BibitemShut {NoStop}%
\bibitem [{\citenamefont {Zache}\ \emph {et~al.}(2017)\citenamefont {Zache},
  \citenamefont {Kasper},\ and\ \citenamefont {Berges}}]{Zache:2017dnz}%
  \BibitemOpen
  \bibfield  {author} {\bibinfo {author} {\bibfnamefont {T.~V.}\ \bibnamefont
  {Zache}}, \bibinfo {author} {\bibfnamefont {V.}~\bibnamefont {Kasper}},\ and\
  \bibinfo {author} {\bibfnamefont {J.}~\bibnamefont {Berges}},\ }\bibfield
  {title} {\bibinfo {title} {{Inflationary preheating dynamics with two-species
  condensates}},\ }\href {https://doi.org/10.1103/PhysRevA.95.063629}
  {\bibfield  {journal} {\bibinfo  {journal} {Phys. Rev. A}\ }\textbf {\bibinfo
  {volume} {95}},\ \bibinfo {pages} {063629} (\bibinfo {year} {2017})},\
  \Eprint {https://arxiv.org/abs/1704.02271} {arXiv:1704.02271
  [cond-mat.quant-gas]} \BibitemShut {NoStop}%
\bibitem [{\citenamefont {Eckel}\ \emph {et~al.}(2018)\citenamefont {Eckel},
  \citenamefont {Kumar}, \citenamefont {Jacobson}, \citenamefont {Spielman},\
  and\ \citenamefont {Campbell}}]{Eckel:2017uqx}%
  \BibitemOpen
  \bibfield  {author} {\bibinfo {author} {\bibfnamefont {S.}~\bibnamefont
  {Eckel}}, \bibinfo {author} {\bibfnamefont {A.}~\bibnamefont {Kumar}},
  \bibinfo {author} {\bibfnamefont {T.}~\bibnamefont {Jacobson}}, \bibinfo
  {author} {\bibfnamefont {I.~B.}\ \bibnamefont {Spielman}},\ and\ \bibinfo
  {author} {\bibfnamefont {G.~K.}\ \bibnamefont {Campbell}},\ }\bibfield
  {title} {\bibinfo {title} {{A rapidly expanding Bose-Einstein condensate: an
  expanding universe in the lab}},\ }\href
  {https://doi.org/10.1103/PhysRevX.8.021021} {\bibfield  {journal} {\bibinfo
  {journal} {Phys. Rev. X}\ }\textbf {\bibinfo {volume} {8}},\ \bibinfo {pages}
  {021021} (\bibinfo {year} {2018})},\ \Eprint
  {https://arxiv.org/abs/1710.05800} {arXiv:1710.05800 [cond-mat.quant-gas]}
  \BibitemShut {NoStop}%
\bibitem [{\citenamefont {Abel}\ and\ \citenamefont
  {Spannowsky}(2021)}]{Abel:2020qzm}%
  \BibitemOpen
  \bibfield  {author} {\bibinfo {author} {\bibfnamefont {S.}~\bibnamefont
  {Abel}}\ and\ \bibinfo {author} {\bibfnamefont {M.}~\bibnamefont
  {Spannowsky}},\ }\bibfield  {title} {\bibinfo {title}
  {{Quantum-Field-Theoretic Simulation Platform for Observing the Fate of the
  False Vacuum}},\ }\href {https://doi.org/10.1103/PRXQuantum.2.010349}
  {\bibfield  {journal} {\bibinfo  {journal} {PRX Quantum}\ }\textbf {\bibinfo
  {volume} {2}},\ \bibinfo {pages} {010349} (\bibinfo {year} {2021})},\ \Eprint
  {https://arxiv.org/abs/2006.06003} {arXiv:2006.06003 [hep-th]} \BibitemShut
  {NoStop}%
\bibitem [{\citenamefont {Chatrchyan}\ \emph {et~al.}(2021)\citenamefont
  {Chatrchyan}, \citenamefont {Geier}, \citenamefont {Oberthaler},
  \citenamefont {Berges},\ and\ \citenamefont {Hauke}}]{Chatrchyan:2020cxs}%
  \BibitemOpen
  \bibfield  {author} {\bibinfo {author} {\bibfnamefont {A.}~\bibnamefont
  {Chatrchyan}}, \bibinfo {author} {\bibfnamefont {K.~T.}\ \bibnamefont
  {Geier}}, \bibinfo {author} {\bibfnamefont {M.~K.}\ \bibnamefont
  {Oberthaler}}, \bibinfo {author} {\bibfnamefont {J.}~\bibnamefont {Berges}},\
  and\ \bibinfo {author} {\bibfnamefont {P.}~\bibnamefont {Hauke}},\ }\bibfield
   {title} {\bibinfo {title} {{Analog cosmological reheating in an ultracold
  Bose gas}},\ }\href {https://doi.org/10.1103/PhysRevA.104.023302} {\bibfield
  {journal} {\bibinfo  {journal} {Phys. Rev. A}\ }\textbf {\bibinfo {volume}
  {104}},\ \bibinfo {pages} {023302} (\bibinfo {year} {2021})},\ \Eprint
  {https://arxiv.org/abs/2008.02290} {arXiv:2008.02290 [cond-mat.quant-gas]}
  \BibitemShut {NoStop}%
\bibitem [{\citenamefont {Milsted}\ \emph {et~al.}(2022)\citenamefont
  {Milsted}, \citenamefont {Liu}, \citenamefont {Preskill},\ and\ \citenamefont
  {Vidal}}]{Milsted:2020jmf}%
  \BibitemOpen
  \bibfield  {author} {\bibinfo {author} {\bibfnamefont {A.}~\bibnamefont
  {Milsted}}, \bibinfo {author} {\bibfnamefont {J.}~\bibnamefont {Liu}},
  \bibinfo {author} {\bibfnamefont {J.}~\bibnamefont {Preskill}},\ and\
  \bibinfo {author} {\bibfnamefont {G.}~\bibnamefont {Vidal}},\ }\bibfield
  {title} {\bibinfo {title} {{Collisions of False-Vacuum Bubble Walls in a
  Quantum Spin Chain}},\ }\href {https://doi.org/10.1103/PRXQuantum.3.020316}
  {\bibfield  {journal} {\bibinfo  {journal} {PRX Quantum}\ }\textbf {\bibinfo
  {volume} {3}},\ \bibinfo {pages} {020316} (\bibinfo {year} {2022})},\ \Eprint
  {https://arxiv.org/abs/2012.07243} {arXiv:2012.07243 [quant-ph]} \BibitemShut
  {NoStop}%
\bibitem [{\citenamefont {Banik}\ \emph {et~al.}(2022)\citenamefont {Banik},
  \citenamefont {Galan}, \citenamefont {Sosa-Martinez}, \citenamefont
  {Anderson}, \citenamefont {Eckel}, \citenamefont {Spielman},\ and\
  \citenamefont {Campbell}}]{Banik:2021xjn}%
  \BibitemOpen
  \bibfield  {author} {\bibinfo {author} {\bibfnamefont {S.}~\bibnamefont
  {Banik}}, \bibinfo {author} {\bibfnamefont {M.~G.}\ \bibnamefont {Galan}},
  \bibinfo {author} {\bibfnamefont {H.}~\bibnamefont {Sosa-Martinez}}, \bibinfo
  {author} {\bibfnamefont {M.}~\bibnamefont {Anderson}}, \bibinfo {author}
  {\bibfnamefont {S.}~\bibnamefont {Eckel}}, \bibinfo {author} {\bibfnamefont
  {I.~B.}\ \bibnamefont {Spielman}},\ and\ \bibinfo {author} {\bibfnamefont
  {G.~K.}\ \bibnamefont {Campbell}},\ }\bibfield  {title} {\bibinfo {title}
  {{Accurate Determination of Hubble Attenuation and Amplification in Expanding
  and Contracting Cold-Atom Universes}},\ }\href
  {https://doi.org/10.1103/PhysRevLett.128.090401} {\bibfield  {journal}
  {\bibinfo  {journal} {Phys. Rev. Lett.}\ }\textbf {\bibinfo {volume} {128}},\
  \bibinfo {pages} {090401} (\bibinfo {year} {2022})},\ \Eprint
  {https://arxiv.org/abs/2107.08097} {arXiv:2107.08097 [quant-ph]} \BibitemShut
  {NoStop}%
\bibitem [{\citenamefont {Lagnese}\ \emph {et~al.}(2021)\citenamefont
  {Lagnese}, \citenamefont {Surace}, \citenamefont {Kormos},\ and\
  \citenamefont {Calabrese}}]{Lagnese:2021grb}%
  \BibitemOpen
  \bibfield  {author} {\bibinfo {author} {\bibfnamefont {G.}~\bibnamefont
  {Lagnese}}, \bibinfo {author} {\bibfnamefont {F.~M.}\ \bibnamefont {Surace}},
  \bibinfo {author} {\bibfnamefont {M.}~\bibnamefont {Kormos}},\ and\ \bibinfo
  {author} {\bibfnamefont {P.}~\bibnamefont {Calabrese}},\ }\bibfield  {title}
  {\bibinfo {title} {{False vacuum decay in quantum spin chains}},\ }\href
  {https://doi.org/10.1103/PhysRevB.104.L201106} {\bibfield  {journal}
  {\bibinfo  {journal} {Phys. Rev. B}\ }\textbf {\bibinfo {volume} {104}},\
  \bibinfo {pages} {L201106} (\bibinfo {year} {2021})},\ \Eprint
  {https://arxiv.org/abs/2107.10176} {arXiv:2107.10176 [cond-mat.stat-mech]}
  \BibitemShut {NoStop}%
\bibitem [{\citenamefont {Viermann}\ \emph {et~al.}(2022)\citenamefont
  {Viermann}, \citenamefont {Sparn}, \citenamefont {Liebster}, \citenamefont
  {Hans}, \citenamefont {Kath}, \citenamefont {Parra-L{\'o}pez}, \citenamefont
  {Tolosa-Sime{\'o}n}, \citenamefont {S{\'a}nchez-Kuntz}, \citenamefont {Haas},
  \citenamefont {Strobel}, \citenamefont {Floerchinger},\ and\ \citenamefont
  {Oberthaler}}]{Viermann:2022wgw}%
  \BibitemOpen
  \bibfield  {author} {\bibinfo {author} {\bibfnamefont {C.}~\bibnamefont
  {Viermann}}, \bibinfo {author} {\bibfnamefont {M.}~\bibnamefont {Sparn}},
  \bibinfo {author} {\bibfnamefont {N.}~\bibnamefont {Liebster}}, \bibinfo
  {author} {\bibfnamefont {M.}~\bibnamefont {Hans}}, \bibinfo {author}
  {\bibfnamefont {E.}~\bibnamefont {Kath}}, \bibinfo {author} {\bibfnamefont
  {{\'A}.}~\bibnamefont {Parra-L{\'o}pez}}, \bibinfo {author} {\bibfnamefont
  {M.}~\bibnamefont {Tolosa-Sime{\'o}n}}, \bibinfo {author} {\bibfnamefont
  {N.}~\bibnamefont {S{\'a}nchez-Kuntz}}, \bibinfo {author} {\bibfnamefont
  {T.}~\bibnamefont {Haas}}, \bibinfo {author} {\bibfnamefont {H.}~\bibnamefont
  {Strobel}}, \bibinfo {author} {\bibfnamefont {S.}~\bibnamefont
  {Floerchinger}},\ and\ \bibinfo {author} {\bibfnamefont {M.~K.}\ \bibnamefont
  {Oberthaler}},\ }\bibfield  {title} {\bibinfo {title} {{Quantum field
  simulator for dynamics in curved spacetime}},\ }\href
  {https://doi.org/10.1038/s41586-022-05313-9} {\bibfield  {journal} {\bibinfo
  {journal} {Nature}\ }\textbf {\bibinfo {volume} {611}},\ \bibinfo {pages}
  {260} (\bibinfo {year} {2022})},\ \Eprint {https://arxiv.org/abs/2202.10399}
  {arXiv:2202.10399 [cond-mat.quant-gas]} \BibitemShut {NoStop}%
\bibitem [{\citenamefont {Tolosa-Sime{\'o}n}\ \emph {et~al.}(2022)\citenamefont
  {Tolosa-Sime{\'o}n}, \citenamefont {Parra-L{\'o}pez}, \citenamefont
  {S{\'a}nchez-Kuntz}, \citenamefont {Haas}, \citenamefont {Viermann},
  \citenamefont {Sparn}, \citenamefont {Liebster}, \citenamefont {Hans},
  \citenamefont {Kath}, \citenamefont {Strobel}, \citenamefont {Oberthaler},\
  and\ \citenamefont {Floerchinger}}]{Tolosa-Simeon:2022umw}%
  \BibitemOpen
  \bibfield  {author} {\bibinfo {author} {\bibfnamefont {M.}~\bibnamefont
  {Tolosa-Sime{\'o}n}}, \bibinfo {author} {\bibfnamefont {{\'A}.}~\bibnamefont
  {Parra-L{\'o}pez}}, \bibinfo {author} {\bibfnamefont {N.}~\bibnamefont
  {S{\'a}nchez-Kuntz}}, \bibinfo {author} {\bibfnamefont {T.}~\bibnamefont
  {Haas}}, \bibinfo {author} {\bibfnamefont {C.}~\bibnamefont {Viermann}},
  \bibinfo {author} {\bibfnamefont {M.}~\bibnamefont {Sparn}}, \bibinfo
  {author} {\bibfnamefont {N.}~\bibnamefont {Liebster}}, \bibinfo {author}
  {\bibfnamefont {M.}~\bibnamefont {Hans}}, \bibinfo {author} {\bibfnamefont
  {E.}~\bibnamefont {Kath}}, \bibinfo {author} {\bibfnamefont {H.}~\bibnamefont
  {Strobel}}, \bibinfo {author} {\bibfnamefont {M.~K.}\ \bibnamefont
  {Oberthaler}},\ and\ \bibinfo {author} {\bibfnamefont {S.}~\bibnamefont
  {Floerchinger}},\ }\bibfield  {title} {\bibinfo {title} {{Curved and
  expanding spacetime geometries in Bose-Einstein condensates}},\ }\href
  {https://doi.org/10.1103/PhysRevA.106.033313} {\bibfield  {journal} {\bibinfo
   {journal} {Phys. Rev. A}\ }\textbf {\bibinfo {volume} {106}},\ \bibinfo
  {pages} {033313} (\bibinfo {year} {2022})},\ \Eprint
  {https://arxiv.org/abs/2202.10441} {arXiv:2202.10441 [cond-mat.quant-gas]}
  \BibitemShut {NoStop}%
\bibitem [{\citenamefont {Tajik}\ \emph {et~al.}(2023)\citenamefont {Tajik},
  \citenamefont {Gluza}, \citenamefont {Sebe}, \citenamefont
  {Sch{\"u}ttelkopf}, \citenamefont {Cataldini}, \citenamefont {Sabino},
  \citenamefont {M{\o}ller}, \citenamefont {Ji}, \citenamefont {Erne},
  \citenamefont {Guarnieri}, \citenamefont {Sotiriadis}, \citenamefont
  {Eisert},\ and\ \citenamefont {Schmiedmayer}}]{Tajik:2022lyt}%
  \BibitemOpen
  \bibfield  {author} {\bibinfo {author} {\bibfnamefont {M.}~\bibnamefont
  {Tajik}}, \bibinfo {author} {\bibfnamefont {M.}~\bibnamefont {Gluza}},
  \bibinfo {author} {\bibfnamefont {N.}~\bibnamefont {Sebe}}, \bibinfo {author}
  {\bibfnamefont {P.}~\bibnamefont {Sch{\"u}ttelkopf}}, \bibinfo {author}
  {\bibfnamefont {F.}~\bibnamefont {Cataldini}}, \bibinfo {author}
  {\bibfnamefont {J.}~\bibnamefont {Sabino}}, \bibinfo {author} {\bibfnamefont
  {F.}~\bibnamefont {M{\o}ller}}, \bibinfo {author} {\bibfnamefont {S.-C.}\
  \bibnamefont {Ji}}, \bibinfo {author} {\bibfnamefont {S.}~\bibnamefont
  {Erne}}, \bibinfo {author} {\bibfnamefont {G.}~\bibnamefont {Guarnieri}},
  \bibinfo {author} {\bibfnamefont {S.}~\bibnamefont {Sotiriadis}}, \bibinfo
  {author} {\bibfnamefont {J.}~\bibnamefont {Eisert}},\ and\ \bibinfo {author}
  {\bibfnamefont {J.}~\bibnamefont {Schmiedmayer}},\ }\bibfield  {title}
  {\bibinfo {title} {{Experimental Observation of Curved Light-Cones in a
  Quantum Field Simulator}},\ }\href {https://doi.org/10.1073/pnas.2301287120}
  {\bibfield  {journal} {\bibinfo  {journal} {Proc. Nat. Acad. Sci.}\ }\textbf
  {\bibinfo {volume} {120}},\ \bibinfo {pages} {e2301287120} (\bibinfo {year}
  {2023})},\ \Eprint {https://arxiv.org/abs/2209.09132} {arXiv:2209.09132
  [cond-mat.quant-gas]} \BibitemShut {NoStop}%
\bibitem [{\citenamefont {Darbha}\ \emph {et~al.}(2024)\citenamefont {Darbha},
  \citenamefont {Kornja{\v{c}}a}, \citenamefont {Liu}, \citenamefont
  {Balewski}, \citenamefont {Hirsbrunner}, \citenamefont {Lopes}, \citenamefont
  {Wang}, \citenamefont {{Van Beeumen}}, \citenamefont {Camps},\ and\
  \citenamefont {Klymko}}]{Darbha:2024srr}%
  \BibitemOpen
  \bibfield  {author} {\bibinfo {author} {\bibfnamefont {S.}~\bibnamefont
  {Darbha}}, \bibinfo {author} {\bibfnamefont {M.}~\bibnamefont
  {Kornja{\v{c}}a}}, \bibinfo {author} {\bibfnamefont {F.}~\bibnamefont {Liu}},
  \bibinfo {author} {\bibfnamefont {J.}~\bibnamefont {Balewski}}, \bibinfo
  {author} {\bibfnamefont {M.~R.}\ \bibnamefont {Hirsbrunner}}, \bibinfo
  {author} {\bibfnamefont {P.~L.~S.}\ \bibnamefont {Lopes}}, \bibinfo {author}
  {\bibfnamefont {S.-T.}\ \bibnamefont {Wang}}, \bibinfo {author}
  {\bibfnamefont {R.}~\bibnamefont {{Van Beeumen}}}, \bibinfo {author}
  {\bibfnamefont {D.}~\bibnamefont {Camps}},\ and\ \bibinfo {author}
  {\bibfnamefont {K.}~\bibnamefont {Klymko}},\ }\bibfield  {title} {\bibinfo
  {title} {{False vacuum decay and nucleation dynamics in neutral atom
  systems}},\ }\href {https://doi.org/10.1103/PhysRevB.110.155103} {\bibfield
  {journal} {\bibinfo  {journal} {Phys. Rev. B}\ }\textbf {\bibinfo {volume}
  {110}},\ \bibinfo {pages} {155103} (\bibinfo {year} {2024})},\ \Eprint
  {https://arxiv.org/abs/2404.12360} {arXiv:2404.12360 [quant-ph]} \BibitemShut
  {NoStop}%
\bibitem [{\citenamefont {Schmidt}\ \emph {et~al.}(2024)\citenamefont
  {Schmidt}, \citenamefont {Parra-L{\'o}pez}, \citenamefont
  {Tolosa-Sime{\'o}n}, \citenamefont {Sparn}, \citenamefont {Kath},
  \citenamefont {Liebster}, \citenamefont {Duchene}, \citenamefont {Strobel},
  \citenamefont {Oberthaler},\ and\ \citenamefont
  {Floerchinger}}]{Schmidt:2024zpg}%
  \BibitemOpen
  \bibfield  {author} {\bibinfo {author} {\bibfnamefont {C.~F.}\ \bibnamefont
  {Schmidt}}, \bibinfo {author} {\bibfnamefont {{\'A}.}~\bibnamefont
  {Parra-L{\'o}pez}}, \bibinfo {author} {\bibfnamefont {M.}~\bibnamefont
  {Tolosa-Sime{\'o}n}}, \bibinfo {author} {\bibfnamefont {M.}~\bibnamefont
  {Sparn}}, \bibinfo {author} {\bibfnamefont {E.}~\bibnamefont {Kath}},
  \bibinfo {author} {\bibfnamefont {N.}~\bibnamefont {Liebster}}, \bibinfo
  {author} {\bibfnamefont {J.}~\bibnamefont {Duchene}}, \bibinfo {author}
  {\bibfnamefont {H.}~\bibnamefont {Strobel}}, \bibinfo {author} {\bibfnamefont
  {M.~K.}\ \bibnamefont {Oberthaler}},\ and\ \bibinfo {author} {\bibfnamefont
  {S.}~\bibnamefont {Floerchinger}},\ }\bibfield  {title} {\bibinfo {title}
  {{Cosmological particle production in a quantum field simulator as a quantum
  mechanical scattering problem}},\ }\href
  {https://doi.org/10.1103/PhysRevD.110.123523} {\bibfield  {journal} {\bibinfo
   {journal} {Phys. Rev. D}\ }\textbf {\bibinfo {volume} {110}},\ \bibinfo
  {pages} {123523} (\bibinfo {year} {2024})},\ \Eprint
  {https://arxiv.org/abs/2406.08094} {arXiv:2406.08094 [gr-qc]} \BibitemShut
  {NoStop}%
\bibitem [{\citenamefont {Zhu}\ \emph {et~al.}(2024)\citenamefont {Zhu},
  \citenamefont {Liu}, \citenamefont {Lagnese}, \citenamefont {Surace},
  \citenamefont {Zhang}, \citenamefont {He}, \citenamefont {Halimeh},
  \citenamefont {Dalmonte}, \citenamefont {Morampudi}, \citenamefont {Wilczek},
  \citenamefont {Yuan},\ and\ \citenamefont {Pan}}]{Zhu:2024dvz}%
  \BibitemOpen
  \bibfield  {author} {\bibinfo {author} {\bibfnamefont {Z.-H.}\ \bibnamefont
  {Zhu}}, \bibinfo {author} {\bibfnamefont {Y.}~\bibnamefont {Liu}}, \bibinfo
  {author} {\bibfnamefont {G.}~\bibnamefont {Lagnese}}, \bibinfo {author}
  {\bibfnamefont {F.~M.}\ \bibnamefont {Surace}}, \bibinfo {author}
  {\bibfnamefont {W.-Y.}\ \bibnamefont {Zhang}}, \bibinfo {author}
  {\bibfnamefont {M.-G.}\ \bibnamefont {He}}, \bibinfo {author} {\bibfnamefont
  {J.~C.}\ \bibnamefont {Halimeh}}, \bibinfo {author} {\bibfnamefont
  {M.}~\bibnamefont {Dalmonte}}, \bibinfo {author} {\bibfnamefont {S.~C.}\
  \bibnamefont {Morampudi}}, \bibinfo {author} {\bibfnamefont {F.}~\bibnamefont
  {Wilczek}}, \bibinfo {author} {\bibfnamefont {Z.-S.}\ \bibnamefont {Yuan}},\
  and\ \bibinfo {author} {\bibfnamefont {J.-W.}\ \bibnamefont {Pan}},\
  }\bibfield  {title} {\bibinfo {title} {{Probing false vacuum decay on a
  cold-atom gauge-theory quantum simulator}},\ }\href@noop {} {\bibfield
  {journal} {\bibinfo  {journal} {{}}\ } (\bibinfo {year} {2024})},\ \Eprint
  {https://arxiv.org/abs/2411.12565} {arXiv:2411.12565 [cond-mat.quant-gas]}
  \BibitemShut {NoStop}%
\bibitem [{\citenamefont {Sch{\"u}tzhold}(2025)}]{Schutzhold:2025qna}%
  \BibitemOpen
  \bibfield  {author} {\bibinfo {author} {\bibfnamefont {R.}~\bibnamefont
  {Sch{\"u}tzhold}},\ }\bibfield  {title} {\bibinfo {title} {{Ultra-cold atoms
  as quantum simulators for relativistic phenomena}},\ }\href@noop {}
  {\bibfield  {journal} {\bibinfo  {journal} {{}}\ } (\bibinfo {year}
  {2025})},\ \Eprint {https://arxiv.org/abs/2501.03785} {arXiv:2501.03785
  [quant-ph]} \BibitemShut {NoStop}%
\bibitem [{\citenamefont {Coleman}(1977)}]{Coleman:1977py}%
  \BibitemOpen
  \bibfield  {author} {\bibinfo {author} {\bibfnamefont {S.~R.}\ \bibnamefont
  {Coleman}},\ }\bibfield  {title} {\bibinfo {title} {{The Fate of the False
  Vacuum. 1. Semiclassical Theory}},\ }\href
  {https://doi.org/10.1103/PhysRevD.16.1248} {\bibfield  {journal} {\bibinfo
  {journal} {Phys. Rev. D}\ }\textbf {\bibinfo {volume} {15}},\ \bibinfo
  {pages} {2929} (\bibinfo {year} {1977})},\ \bibinfo {note} {[Erratum: Phys.
  Rev. D 16, 1248(E) (1977)]}\BibitemShut {NoStop}%
\bibitem [{\citenamefont {Callan}\ and\ \citenamefont
  {Coleman}(1977)}]{Callan:1977pt}%
  \BibitemOpen
  \bibfield  {author} {\bibinfo {author} {\bibfnamefont {C.~G.}\ \bibnamefont
  {Callan}, \bibfnamefont {Jr.}}\ and\ \bibinfo {author} {\bibfnamefont
  {S.~R.}\ \bibnamefont {Coleman}},\ }\bibfield  {title} {\bibinfo {title}
  {{The Fate of the False Vacuum. 2. First Quantum Corrections}},\ }\href
  {https://doi.org/10.1103/PhysRevD.16.1762} {\bibfield  {journal} {\bibinfo
  {journal} {Phys. Rev. D}\ }\textbf {\bibinfo {volume} {16}},\ \bibinfo
  {pages} {1762} (\bibinfo {year} {1977})}\BibitemShut {NoStop}%
\bibitem [{\citenamefont {Pirvu}\ \emph {et~al.}(2022)\citenamefont {Pirvu},
  \citenamefont {Braden},\ and\ \citenamefont {Johnson}}]{Pirvu:2021roq}%
  \BibitemOpen
  \bibfield  {author} {\bibinfo {author} {\bibfnamefont {D.}~\bibnamefont
  {Pirvu}}, \bibinfo {author} {\bibfnamefont {J.}~\bibnamefont {Braden}},\ and\
  \bibinfo {author} {\bibfnamefont {M.~C.}\ \bibnamefont {Johnson}},\
  }\bibfield  {title} {\bibinfo {title} {{Bubble clustering in cosmological
  first order phase transitions}},\ }\href
  {https://doi.org/10.1103/PhysRevD.105.043510} {\bibfield  {journal} {\bibinfo
   {journal} {Phys. Rev. D}\ }\textbf {\bibinfo {volume} {105}},\ \bibinfo
  {pages} {043510} (\bibinfo {year} {2022})},\ \Eprint
  {https://arxiv.org/abs/2109.04496} {arXiv:2109.04496 [hep-th]} \BibitemShut
  {NoStop}%
\bibitem [{\citenamefont {De~Luca}\ \emph {et~al.}(2021)\citenamefont
  {De~Luca}, \citenamefont {Franciolini},\ and\ \citenamefont
  {Riotto}}]{DeLuca:2021mlh}%
  \BibitemOpen
  \bibfield  {author} {\bibinfo {author} {\bibfnamefont {V.}~\bibnamefont
  {De~Luca}}, \bibinfo {author} {\bibfnamefont {G.}~\bibnamefont
  {Franciolini}},\ and\ \bibinfo {author} {\bibfnamefont {A.}~\bibnamefont
  {Riotto}},\ }\bibfield  {title} {\bibinfo {title} {{Bubble correlation in
  first-order phase transitions}},\ }\href
  {https://doi.org/10.1103/PhysRevD.104.123539} {\bibfield  {journal} {\bibinfo
   {journal} {Phys. Rev. D}\ }\textbf {\bibinfo {volume} {104}},\ \bibinfo
  {pages} {123539} (\bibinfo {year} {2021})},\ \Eprint
  {https://arxiv.org/abs/2110.04229} {arXiv:2110.04229 [hep-ph]} \BibitemShut
  {NoStop}%
\bibitem [{\citenamefont {P\^\i{}rvu}\ \emph {et~al.}(2024)\citenamefont
  {P\^\i{}rvu}, \citenamefont {Johnson},\ and\ \citenamefont
  {Sibiryakov}}]{Pirvu:2023plk}%
  \BibitemOpen
  \bibfield  {author} {\bibinfo {author} {\bibfnamefont {D.}~\bibnamefont
  {P\^\i{}rvu}}, \bibinfo {author} {\bibfnamefont {M.~C.}\ \bibnamefont
  {Johnson}},\ and\ \bibinfo {author} {\bibfnamefont {S.}~\bibnamefont
  {Sibiryakov}},\ }\bibfield  {title} {\bibinfo {title} {{Bubble velocities and
  oscillon precursors in first-order phase transitions}},\ }\href
  {https://doi.org/10.1007/JHEP11(2024)064} {\bibfield  {journal} {\bibinfo
  {journal} {JHEP}\ }\textbf {\bibinfo {volume} {11}}\bibfield  {number}
  {\bibinfo  {number} { (2024)},\ \bibinfo {pages} {064}},\ }\Eprint
  {https://arxiv.org/abs/2312.13364} {arXiv:2312.13364 [hep-th]} \BibitemShut
  {NoStop}%
\bibitem [{\citenamefont {Batini}\ \emph {et~al.}(2024)\citenamefont {Batini},
  \citenamefont {Chatrchyan},\ and\ \citenamefont {Berges}}]{Batini:2023zpi}%
  \BibitemOpen
  \bibfield  {author} {\bibinfo {author} {\bibfnamefont {L.}~\bibnamefont
  {Batini}}, \bibinfo {author} {\bibfnamefont {A.}~\bibnamefont {Chatrchyan}},\
  and\ \bibinfo {author} {\bibfnamefont {J.}~\bibnamefont {Berges}},\
  }\bibfield  {title} {\bibinfo {title} {{Real-time dynamics of false vacuum
  decay}},\ }\href {https://doi.org/10.1103/PhysRevD.109.023502} {\bibfield
  {journal} {\bibinfo  {journal} {Phys. Rev. D}\ }\textbf {\bibinfo {volume}
  {109}},\ \bibinfo {pages} {023502} (\bibinfo {year} {2024})},\ \Eprint
  {https://arxiv.org/abs/2310.04206} {arXiv:2310.04206 [hep-th]} \BibitemShut
  {NoStop}%
\bibitem [{\citenamefont {P{\^i}rvu}\ \emph {et~al.}(2024)\citenamefont
  {P{\^i}rvu}, \citenamefont {Shkerin},\ and\ \citenamefont
  {Sibiryakov}}]{Pirvu:2024ova}%
  \BibitemOpen
  \bibfield  {author} {\bibinfo {author} {\bibfnamefont {D.}~\bibnamefont
  {P{\^i}rvu}}, \bibinfo {author} {\bibfnamefont {A.}~\bibnamefont {Shkerin}},\
  and\ \bibinfo {author} {\bibfnamefont {S.}~\bibnamefont {Sibiryakov}},\
  }\bibfield  {title} {\bibinfo {title} {{Thermal False Vacuum Decay Is Not
  What It Seems}},\ }\href@noop {} {\bibfield  {journal} {\bibinfo  {journal}
  {{}}\ } (\bibinfo {year} {2024})},\ \Eprint
  {https://arxiv.org/abs/2407.06263} {arXiv:2407.06263 [hep-th]} \BibitemShut
  {NoStop}%
\bibitem [{\citenamefont {P\^\i{}rvu}\ \emph {et~al.}(2024)\citenamefont
  {P\^\i{}rvu}, \citenamefont {Shkerin},\ and\ \citenamefont
  {Sibiryakov}}]{Pirvu:2024nbe}%
  \BibitemOpen
  \bibfield  {author} {\bibinfo {author} {\bibfnamefont {D.}~\bibnamefont
  {P\^\i{}rvu}}, \bibinfo {author} {\bibfnamefont {A.}~\bibnamefont
  {Shkerin}},\ and\ \bibinfo {author} {\bibfnamefont {S.}~\bibnamefont
  {Sibiryakov}},\ }\bibfield  {title} {\bibinfo {title} {{Thermal false vacuum
  decay in (1+1) dimensions: Evidence for nonequilibrium dynamics}},\ }\href
  {https://doi.org/10.1142/S0217751X24450076} {\bibfield  {journal} {\bibinfo
  {journal} {Int. J. Mod. Phys. A}\ }\textbf {\bibinfo {volume} {39}},\
  \bibinfo {pages} {2445007} (\bibinfo {year} {2024})},\ \Eprint
  {https://arxiv.org/abs/2408.06411} {arXiv:2408.06411 [hep-th]} \BibitemShut
  {NoStop}%
\bibitem [{\citenamefont {Guth}(2007)}]{Guth:2007ng}%
  \BibitemOpen
  \bibfield  {author} {\bibinfo {author} {\bibfnamefont {A.~H.}\ \bibnamefont
  {Guth}},\ }\bibfield  {title} {\bibinfo {title} {{Eternal inflation and its
  implications}},\ }\href {https://doi.org/10.1088/1751-8113/40/25/S25}
  {\bibfield  {journal} {\bibinfo  {journal} {J. Phys. A}\ }\textbf {\bibinfo
  {volume} {40}},\ \bibinfo {pages} {6811} (\bibinfo {year} {2007})},\ \Eprint
  {https://arxiv.org/abs/hep-th/0702178} {arXiv:hep-th/0702178} \BibitemShut
  {NoStop}%
\bibitem [{\citenamefont {Aguirre}\ \emph {et~al.}(2007)\citenamefont
  {Aguirre}, \citenamefont {Johnson},\ and\ \citenamefont
  {Shomer}}]{Aguirre:2007an}%
  \BibitemOpen
  \bibfield  {author} {\bibinfo {author} {\bibfnamefont {A.}~\bibnamefont
  {Aguirre}}, \bibinfo {author} {\bibfnamefont {M.~C.}\ \bibnamefont
  {Johnson}},\ and\ \bibinfo {author} {\bibfnamefont {A.}~\bibnamefont
  {Shomer}},\ }\bibfield  {title} {\bibinfo {title} {{Towards observable
  signatures of other bubble universes}},\ }\href
  {https://doi.org/10.1103/PhysRevD.76.063509} {\bibfield  {journal} {\bibinfo
  {journal} {Phys. Rev. D}\ }\textbf {\bibinfo {volume} {76}},\ \bibinfo
  {pages} {063509} (\bibinfo {year} {2007})},\ \Eprint
  {https://arxiv.org/abs/0704.3473} {arXiv:0704.3473 [hep-th]} \BibitemShut
  {NoStop}%
\bibitem [{\citenamefont {Aguirre}(2007)}]{Aguirre:2007gy}%
  \BibitemOpen
  \bibfield  {author} {\bibinfo {author} {\bibfnamefont {A.}~\bibnamefont
  {Aguirre}},\ }\bibfield  {title} {\bibinfo {title} {{Eternal Inflation, past
  and future}},\ }\href@noop {} {\bibfield  {journal} {\bibinfo  {journal}
  {{}}\ } (\bibinfo {year} {2007})},\ \Eprint {https://arxiv.org/abs/0712.0571}
  {arXiv:0712.0571 [hep-th]} \BibitemShut {NoStop}%
\bibitem [{\citenamefont {Feeney}\ \emph
  {et~al.}(2011{\natexlab{a}})\citenamefont {Feeney}, \citenamefont {Johnson},
  \citenamefont {Mortlock},\ and\ \citenamefont {Peiris}}]{Feeney:2010jj}%
  \BibitemOpen
  \bibfield  {author} {\bibinfo {author} {\bibfnamefont {S.~M.}\ \bibnamefont
  {Feeney}}, \bibinfo {author} {\bibfnamefont {M.~C.}\ \bibnamefont {Johnson}},
  \bibinfo {author} {\bibfnamefont {D.~J.}\ \bibnamefont {Mortlock}},\ and\
  \bibinfo {author} {\bibfnamefont {H.~V.}\ \bibnamefont {Peiris}},\ }\bibfield
   {title} {\bibinfo {title} {{First Observational Tests of Eternal
  Inflation}},\ }\href {https://doi.org/10.1103/PhysRevLett.107.071301}
  {\bibfield  {journal} {\bibinfo  {journal} {Phys. Rev. Lett.}\ }\textbf
  {\bibinfo {volume} {107}},\ \bibinfo {pages} {071301} (\bibinfo {year}
  {2011}{\natexlab{a}})},\ \Eprint {https://arxiv.org/abs/1012.1995}
  {arXiv:1012.1995 [astro-ph.CO]} \BibitemShut {NoStop}%
\bibitem [{\citenamefont {Feeney}\ \emph
  {et~al.}(2011{\natexlab{b}})\citenamefont {Feeney}, \citenamefont {Johnson},
  \citenamefont {Mortlock},\ and\ \citenamefont {Peiris}}]{Feeney:2010dd}%
  \BibitemOpen
  \bibfield  {author} {\bibinfo {author} {\bibfnamefont {S.~M.}\ \bibnamefont
  {Feeney}}, \bibinfo {author} {\bibfnamefont {M.~C.}\ \bibnamefont {Johnson}},
  \bibinfo {author} {\bibfnamefont {D.~J.}\ \bibnamefont {Mortlock}},\ and\
  \bibinfo {author} {\bibfnamefont {H.~V.}\ \bibnamefont {Peiris}},\ }\bibfield
   {title} {\bibinfo {title} {{First Observational Tests of Eternal Inflation:
  Analysis Methods and WMAP 7-Year Results}},\ }\href
  {https://doi.org/10.1103/PhysRevD.84.043507} {\bibfield  {journal} {\bibinfo
  {journal} {Phys. Rev. D}\ }\textbf {\bibinfo {volume} {84}},\ \bibinfo
  {pages} {043507} (\bibinfo {year} {2011}{\natexlab{b}})},\ \Eprint
  {https://arxiv.org/abs/1012.3667} {arXiv:1012.3667 [astro-ph.CO]}
  \BibitemShut {NoStop}%
\bibitem [{\citenamefont {Kuzmin}\ \emph {et~al.}(1985)\citenamefont {Kuzmin},
  \citenamefont {Rubakov},\ and\ \citenamefont {Shaposhnikov}}]{Kuzmin:1985mm}%
  \BibitemOpen
  \bibfield  {author} {\bibinfo {author} {\bibfnamefont {V.~A.}\ \bibnamefont
  {Kuzmin}}, \bibinfo {author} {\bibfnamefont {V.~A.}\ \bibnamefont
  {Rubakov}},\ and\ \bibinfo {author} {\bibfnamefont {M.~E.}\ \bibnamefont
  {Shaposhnikov}},\ }\bibfield  {title} {\bibinfo {title} {{On the Anomalous
  Electroweak Baryon Number Nonconservation in the Early Universe}},\ }\href
  {https://doi.org/10.1016/0370-2693(85)91028-7} {\bibfield  {journal}
  {\bibinfo  {journal} {Phys. Lett. B}\ }\textbf {\bibinfo {volume} {155}},\
  \bibinfo {pages} {36} (\bibinfo {year} {1985})}\BibitemShut {NoStop}%
\bibitem [{\citenamefont {Cohen}\ \emph {et~al.}(1993)\citenamefont {Cohen},
  \citenamefont {Kaplan},\ and\ \citenamefont {Nelson}}]{Cohen:1993nk}%
  \BibitemOpen
  \bibfield  {author} {\bibinfo {author} {\bibfnamefont {A.~G.}\ \bibnamefont
  {Cohen}}, \bibinfo {author} {\bibfnamefont {D.~B.}\ \bibnamefont {Kaplan}},\
  and\ \bibinfo {author} {\bibfnamefont {A.~E.}\ \bibnamefont {Nelson}},\
  }\bibfield  {title} {\bibinfo {title} {{Progress in electroweak
  baryogenesis}},\ }\href {https://doi.org/10.1146/annurev.ns.43.120193.000331}
  {\bibfield  {journal} {\bibinfo  {journal} {Ann. Rev. Nucl. Part. Sci.}\
  }\textbf {\bibinfo {volume} {43}},\ \bibinfo {pages} {27} (\bibinfo {year}
  {1993})},\ \Eprint {https://arxiv.org/abs/hep-ph/9302210}
  {arXiv:hep-ph/9302210} \BibitemShut {NoStop}%
\bibitem [{\citenamefont {Morrissey}\ and\ \citenamefont
  {Ramsey-Musolf}(2012)}]{Morrissey:2012db}%
  \BibitemOpen
  \bibfield  {author} {\bibinfo {author} {\bibfnamefont {D.~E.}\ \bibnamefont
  {Morrissey}}\ and\ \bibinfo {author} {\bibfnamefont {M.~J.}\ \bibnamefont
  {Ramsey-Musolf}},\ }\bibfield  {title} {\bibinfo {title} {{Electroweak
  baryogenesis}},\ }\href {https://doi.org/10.1088/1367-2630/14/12/125003}
  {\bibfield  {journal} {\bibinfo  {journal} {New J. Phys.}\ }\textbf {\bibinfo
  {volume} {14}},\ \bibinfo {pages} {125003} (\bibinfo {year} {2012})},\
  \Eprint {https://arxiv.org/abs/1206.2942} {arXiv:1206.2942 [hep-ph]}
  \BibitemShut {NoStop}%
\bibitem [{\citenamefont {Kosowsky}\ \emph {et~al.}(1992)\citenamefont
  {Kosowsky}, \citenamefont {Turner},\ and\ \citenamefont
  {Watkins}}]{Kosowsky:1991ua}%
  \BibitemOpen
  \bibfield  {author} {\bibinfo {author} {\bibfnamefont {A.}~\bibnamefont
  {Kosowsky}}, \bibinfo {author} {\bibfnamefont {M.~S.}\ \bibnamefont
  {Turner}},\ and\ \bibinfo {author} {\bibfnamefont {R.}~\bibnamefont
  {Watkins}},\ }\bibfield  {title} {\bibinfo {title} {{Gravitational radiation
  from colliding vacuum bubbles}},\ }\href
  {https://doi.org/10.1103/PhysRevD.45.4514} {\bibfield  {journal} {\bibinfo
  {journal} {Phys. Rev. D}\ }\textbf {\bibinfo {volume} {45}},\ \bibinfo
  {pages} {4514} (\bibinfo {year} {1992})}\BibitemShut {NoStop}%
\bibitem [{\citenamefont {Kamionkowski}\ \emph {et~al.}(1994)\citenamefont
  {Kamionkowski}, \citenamefont {Kosowsky},\ and\ \citenamefont
  {Turner}}]{Kamionkowski:1993fg}%
  \BibitemOpen
  \bibfield  {author} {\bibinfo {author} {\bibfnamefont {M.}~\bibnamefont
  {Kamionkowski}}, \bibinfo {author} {\bibfnamefont {A.}~\bibnamefont
  {Kosowsky}},\ and\ \bibinfo {author} {\bibfnamefont {M.~S.}\ \bibnamefont
  {Turner}},\ }\bibfield  {title} {\bibinfo {title} {{Gravitational radiation
  from first order phase transitions}},\ }\href
  {https://doi.org/10.1103/PhysRevD.49.2837} {\bibfield  {journal} {\bibinfo
  {journal} {Phys. Rev. D}\ }\textbf {\bibinfo {volume} {49}},\ \bibinfo
  {pages} {2837} (\bibinfo {year} {1994})},\ \Eprint
  {https://arxiv.org/abs/astro-ph/9310044} {arXiv:astro-ph/9310044}
  \BibitemShut {NoStop}%
\bibitem [{\citenamefont {Caprini}\ \emph {et~al.}(2016)\citenamefont
  {Caprini}, \citenamefont {Hindmarsh}, \citenamefont {Huber}, \citenamefont
  {Konstandin}, \citenamefont {Kozaczuk}, \citenamefont {Nardini},
  \citenamefont {No}, \citenamefont {Petiteau}, \citenamefont {Schwaller},
  \citenamefont {Servant},\ and\ \citenamefont {Weir}}]{Caprini:2015zlo}%
  \BibitemOpen
  \bibfield  {author} {\bibinfo {author} {\bibfnamefont {C.}~\bibnamefont
  {Caprini}}, \bibinfo {author} {\bibfnamefont {M.}~\bibnamefont {Hindmarsh}},
  \bibinfo {author} {\bibfnamefont {S.}~\bibnamefont {Huber}}, \bibinfo
  {author} {\bibfnamefont {T.}~\bibnamefont {Konstandin}}, \bibinfo {author}
  {\bibfnamefont {J.}~\bibnamefont {Kozaczuk}}, \bibinfo {author}
  {\bibfnamefont {G.}~\bibnamefont {Nardini}}, \bibinfo {author} {\bibfnamefont
  {J.~M.}\ \bibnamefont {No}}, \bibinfo {author} {\bibfnamefont
  {A.}~\bibnamefont {Petiteau}}, \bibinfo {author} {\bibfnamefont
  {P.}~\bibnamefont {Schwaller}}, \bibinfo {author} {\bibfnamefont
  {G.}~\bibnamefont {Servant}},\ and\ \bibinfo {author} {\bibfnamefont {D.~J.}\
  \bibnamefont {Weir}},\ }\bibfield  {title} {\bibinfo {title} {{Science with
  the space-based interferometer eLISA. II: Gravitational waves from
  cosmological phase transitions}},\ }\href
  {https://doi.org/10.1088/1475-7516/2016/04/001} {\bibfield  {journal}
  {\bibinfo  {journal} {JCAP}\ }\textbf {\bibinfo {volume} {04}}\bibfield
  {number} {\bibinfo  {number} { (2016)},\ \bibinfo {pages} {001}},\ }\Eprint
  {https://arxiv.org/abs/1512.06239} {arXiv:1512.06239 [astro-ph.CO]}
  \BibitemShut {NoStop}%
\bibitem [{\citenamefont {Ellis}\ \emph {et~al.}(2009)\citenamefont {Ellis},
  \citenamefont {Espinosa}, \citenamefont {Giudice}, \citenamefont {Hoecker},\
  and\ \citenamefont {Riotto}}]{Ellis:2009tp}%
  \BibitemOpen
  \bibfield  {author} {\bibinfo {author} {\bibfnamefont {J.}~\bibnamefont
  {Ellis}}, \bibinfo {author} {\bibfnamefont {J.~R.}\ \bibnamefont {Espinosa}},
  \bibinfo {author} {\bibfnamefont {G.~F.}\ \bibnamefont {Giudice}}, \bibinfo
  {author} {\bibfnamefont {A.}~\bibnamefont {Hoecker}},\ and\ \bibinfo {author}
  {\bibfnamefont {A.}~\bibnamefont {Riotto}},\ }\bibfield  {title} {\bibinfo
  {title} {{The Probable Fate of the Standard Model}},\ }\href
  {https://doi.org/10.1016/j.physletb.2009.07.054} {\bibfield  {journal}
  {\bibinfo  {journal} {Phys. Lett. B}\ }\textbf {\bibinfo {volume} {679}},\
  \bibinfo {pages} {369} (\bibinfo {year} {2009})},\ \Eprint
  {https://arxiv.org/abs/0906.0954} {arXiv:0906.0954 [hep-ph]} \BibitemShut
  {NoStop}%
\bibitem [{\citenamefont {Degrassi}\ \emph {et~al.}(2012)\citenamefont
  {Degrassi}, \citenamefont {Di~Vita}, \citenamefont {Elias-Miro},
  \citenamefont {Espinosa}, \citenamefont {Giudice}, \citenamefont {Isidori},\
  and\ \citenamefont {Strumia}}]{Degrassi:2012ry}%
  \BibitemOpen
  \bibfield  {author} {\bibinfo {author} {\bibfnamefont {G.}~\bibnamefont
  {Degrassi}}, \bibinfo {author} {\bibfnamefont {S.}~\bibnamefont {Di~Vita}},
  \bibinfo {author} {\bibfnamefont {J.}~\bibnamefont {Elias-Miro}}, \bibinfo
  {author} {\bibfnamefont {J.~R.}\ \bibnamefont {Espinosa}}, \bibinfo {author}
  {\bibfnamefont {G.~F.}\ \bibnamefont {Giudice}}, \bibinfo {author}
  {\bibfnamefont {G.}~\bibnamefont {Isidori}},\ and\ \bibinfo {author}
  {\bibfnamefont {A.}~\bibnamefont {Strumia}},\ }\bibfield  {title} {\bibinfo
  {title} {{Higgs mass and vacuum stability in the Standard Model at NNLO}},\
  }\href {https://doi.org/10.1007/JHEP08(2012)098} {\bibfield  {journal}
  {\bibinfo  {journal} {JHEP}\ }\textbf {\bibinfo {volume} {08}}\bibfield
  {number} {\bibinfo  {number} { (2012)},\ \bibinfo {pages} {098}},\ }\Eprint
  {https://arxiv.org/abs/1205.6497} {arXiv:1205.6497 [hep-ph]} \BibitemShut
  {NoStop}%
\bibitem [{\citenamefont {Buttazzo}\ \emph {et~al.}(2013)\citenamefont
  {Buttazzo}, \citenamefont {Degrassi}, \citenamefont {Giardino}, \citenamefont
  {Giudice}, \citenamefont {Sala}, \citenamefont {Salvio},\ and\ \citenamefont
  {Strumia}}]{Buttazzo:2013uya}%
  \BibitemOpen
  \bibfield  {author} {\bibinfo {author} {\bibfnamefont {D.}~\bibnamefont
  {Buttazzo}}, \bibinfo {author} {\bibfnamefont {G.}~\bibnamefont {Degrassi}},
  \bibinfo {author} {\bibfnamefont {P.~P.}\ \bibnamefont {Giardino}}, \bibinfo
  {author} {\bibfnamefont {G.~F.}\ \bibnamefont {Giudice}}, \bibinfo {author}
  {\bibfnamefont {F.}~\bibnamefont {Sala}}, \bibinfo {author} {\bibfnamefont
  {A.}~\bibnamefont {Salvio}},\ and\ \bibinfo {author} {\bibfnamefont
  {A.}~\bibnamefont {Strumia}},\ }\bibfield  {title} {\bibinfo {title}
  {{Investigating the near-criticality of the Higgs boson}},\ }\href
  {https://doi.org/10.1007/JHEP12(2013)089} {\bibfield  {journal} {\bibinfo
  {journal} {JHEP}\ }\textbf {\bibinfo {volume} {12}}\bibfield  {number}
  {\bibinfo  {number} { (2013)},\ \bibinfo {pages} {089}},\ }\Eprint
  {https://arxiv.org/abs/1307.3536} {arXiv:1307.3536 [hep-ph]} \BibitemShut
  {NoStop}%
\bibitem [{\citenamefont {Gaunt}\ \emph {et~al.}(2013)\citenamefont {Gaunt},
  \citenamefont {Schmidutz}, \citenamefont {Gotlibovych}, \citenamefont
  {Smith},\ and\ \citenamefont {Hadzibabic}}]{Gaunt:2013box}%
  \BibitemOpen
  \bibfield  {author} {\bibinfo {author} {\bibfnamefont {A.~L.}\ \bibnamefont
  {Gaunt}}, \bibinfo {author} {\bibfnamefont {T.~F.}\ \bibnamefont
  {Schmidutz}}, \bibinfo {author} {\bibfnamefont {I.}~\bibnamefont
  {Gotlibovych}}, \bibinfo {author} {\bibfnamefont {R.~P.}\ \bibnamefont
  {Smith}},\ and\ \bibinfo {author} {\bibfnamefont {Z.}~\bibnamefont
  {Hadzibabic}},\ }\bibfield  {title} {\bibinfo {title} {{Bose-Einstein
  Condensation of Atoms in a Uniform Potential}},\ }\href
  {https://doi.org/10.1103/PhysRevLett.110.200406} {\bibfield  {journal}
  {\bibinfo  {journal} {Phys. Rev. Lett.}\ }\textbf {\bibinfo {volume} {110}},\
  \bibinfo {pages} {200406} (\bibinfo {year} {2013})},\ \Eprint
  {https://arxiv.org/abs/1212.4453} {arXiv:1212.4453 [cond-mat.quant-gas]}
  \BibitemShut {NoStop}%
\bibitem [{\citenamefont {Navon}\ \emph {et~al.}(2021)\citenamefont {Navon},
  \citenamefont {Smith},\ and\ \citenamefont {Hadzibabic}}]{Navon:2021mcf}%
  \BibitemOpen
  \bibfield  {author} {\bibinfo {author} {\bibfnamefont {N.}~\bibnamefont
  {Navon}}, \bibinfo {author} {\bibfnamefont {R.~P.}\ \bibnamefont {Smith}},\
  and\ \bibinfo {author} {\bibfnamefont {Z.}~\bibnamefont {Hadzibabic}},\
  }\bibfield  {title} {\bibinfo {title} {{Quantum gases in optical boxes}},\
  }\href {https://doi.org/10.1038/s41567-021-01403-z} {\bibfield  {journal}
  {\bibinfo  {journal} {Nature Phys.}\ }\textbf {\bibinfo {volume} {17}},\
  \bibinfo {pages} {1334} (\bibinfo {year} {2021})},\ \Eprint
  {https://arxiv.org/abs/2106.09716} {arXiv:2106.09716 [cond-mat.quant-gas]}
  \BibitemShut {NoStop}%
\bibitem [{\citenamefont {Moss}(1985)}]{Moss:1984zf}%
  \BibitemOpen
  \bibfield  {author} {\bibinfo {author} {\bibfnamefont {I.~G.}\ \bibnamefont
  {Moss}},\ }\bibfield  {title} {\bibinfo {title} {{Black hole bubbles}},\
  }\href {https://doi.org/10.1103/PhysRevD.32.1333} {\bibfield  {journal}
  {\bibinfo  {journal} {Phys. Rev. D}\ }\textbf {\bibinfo {volume} {32}},\
  \bibinfo {pages} {1333} (\bibinfo {year} {1985})}\BibitemShut {NoStop}%
\bibitem [{\citenamefont {Gregory}\ \emph {et~al.}(2014)\citenamefont
  {Gregory}, \citenamefont {Moss},\ and\ \citenamefont
  {Withers}}]{Gregory:2013hja}%
  \BibitemOpen
  \bibfield  {author} {\bibinfo {author} {\bibfnamefont {R.}~\bibnamefont
  {Gregory}}, \bibinfo {author} {\bibfnamefont {I.~G.}\ \bibnamefont {Moss}},\
  and\ \bibinfo {author} {\bibfnamefont {B.}~\bibnamefont {Withers}},\
  }\bibfield  {title} {\bibinfo {title} {{Black holes as bubble nucleation
  sites}},\ }\href {https://doi.org/10.1007/JHEP03(2014)081} {\bibfield
  {journal} {\bibinfo  {journal} {JHEP}\ }\textbf {\bibinfo {volume}
  {03}}\bibfield  {number} {\bibinfo  {number} { (2014)},\ \bibinfo {pages}
  {081}},\ }\Eprint {https://arxiv.org/abs/1401.0017} {arXiv:1401.0017
  [hep-th]} \BibitemShut {NoStop}%
\bibitem [{\citenamefont {Caneletti}\ and\ \citenamefont
  {Moss}(2024)}]{Caneletti:2024kww}%
  \BibitemOpen
  \bibfield  {author} {\bibinfo {author} {\bibfnamefont {M.}~\bibnamefont
  {Caneletti}}\ and\ \bibinfo {author} {\bibfnamefont {I.~G.}\ \bibnamefont
  {Moss}},\ }\bibfield  {title} {\bibinfo {title} {{Seeding the decay of the
  false vacuum}},\ }\href {https://doi.org/10.1103/PhysRevD.110.105015}
  {\bibfield  {journal} {\bibinfo  {journal} {Phys. Rev. D}\ }\textbf {\bibinfo
  {volume} {110}},\ \bibinfo {pages} {105015} (\bibinfo {year} {2024})},\
  \Eprint {https://arxiv.org/abs/2408.12229} {arXiv:2408.12229 [hep-th]}
  \BibitemShut {NoStop}%
\bibitem [{\citenamefont {Brown}\ \emph {et~al.}(2025)\citenamefont {Brown},
  \citenamefont {Moss},\ and\ \citenamefont {Billam}}]{Brown:2025wxy}%
  \BibitemOpen
  \bibfield  {author} {\bibinfo {author} {\bibfnamefont {K.}~\bibnamefont
  {Brown}}, \bibinfo {author} {\bibfnamefont {I.~G.}\ \bibnamefont {Moss}},\
  and\ \bibinfo {author} {\bibfnamefont {T.~P.}\ \bibnamefont {Billam}},\
  }\bibfield  {title} {\bibinfo {title} {{Mitigating boundary effects in finite
  temperature simulations of false vacuum decay}},\ }\href@noop {} {\bibfield
  {journal} {\bibinfo  {journal} {{}}\ } (\bibinfo {year} {2025})},\ \Eprint
  {https://arxiv.org/abs/2504.03509} {arXiv:2504.03509 [cond-mat.quant-gas]}
  \BibitemShut {NoStop}%
\bibitem [{\citenamefont {Kashchiev}(2000)}]{Kashchiev:2000nuc}%
  \BibitemOpen
  \bibfield  {author} {\bibinfo {author} {\bibfnamefont {D.}~\bibnamefont
  {Kashchiev}},\ }\href@noop {} {\emph {\bibinfo {title} {{Nucleation}}}}\
  (\bibinfo  {publisher} {Elsevier},\ \bibinfo {year} {2000})\BibitemShut
  {NoStop}%
\bibitem [{\citenamefont {Gallo}\ \emph {et~al.}(2021)\citenamefont {Gallo},
  \citenamefont {Magaletti},\ and\ \citenamefont {Casciola}}]{Gallo:2021het}%
  \BibitemOpen
  \bibfield  {author} {\bibinfo {author} {\bibfnamefont {M.}~\bibnamefont
  {Gallo}}, \bibinfo {author} {\bibfnamefont {F.}~\bibnamefont {Magaletti}},\
  and\ \bibinfo {author} {\bibfnamefont {C.~M.}\ \bibnamefont {Casciola}},\
  }\bibfield  {title} {\bibinfo {title} {{Heterogeneous bubble nucleation
  dynamics}},\ }\href {https://doi.org/10.1017/jfm.2020.761} {\bibfield
  {journal} {\bibinfo  {journal} {J. Fluid Mech.}\ }\textbf {\bibinfo {volume}
  {906}},\ \bibinfo {pages} {A20} (\bibinfo {year} {2021})}\BibitemShut
  {NoStop}%
\bibitem [{\citenamefont {Gauthier}\ \emph {et~al.}(2016)\citenamefont
  {Gauthier}, \citenamefont {Lenton}, \citenamefont {McKay~Parry},
  \citenamefont {Baker}, \citenamefont {Davis}, \citenamefont
  {Rubinsztein-Dunlop},\ and\ \citenamefont {W.}}]{Gauthier:2016dmd}%
  \BibitemOpen
  \bibfield  {author} {\bibinfo {author} {\bibfnamefont {G.}~\bibnamefont
  {Gauthier}}, \bibinfo {author} {\bibfnamefont {I.}~\bibnamefont {Lenton}},
  \bibinfo {author} {\bibfnamefont {N.}~\bibnamefont {McKay~Parry}}, \bibinfo
  {author} {\bibfnamefont {M.}~\bibnamefont {Baker}}, \bibinfo {author}
  {\bibfnamefont {M.~J.}\ \bibnamefont {Davis}}, \bibinfo {author}
  {\bibfnamefont {H.}~\bibnamefont {Rubinsztein-Dunlop}},\ and\ \bibinfo
  {author} {\bibfnamefont {N.~T.}\ \bibnamefont {W.}},\ }\bibfield  {title}
  {\bibinfo {title} {{Direct imaging of a digital-micromirror device for
  configurable microscopic optical potentials}},\ }\href
  {https://doi.org/https://doi.org/10.1364/OPTICA.3.001136} {\bibfield
  {journal} {\bibinfo  {journal} {Optica}\ }\textbf {\bibinfo {volume} {3}},\
  \bibinfo {pages} {1136} (\bibinfo {year} {2016})},\ \Eprint
  {https://arxiv.org/abs/1605.04928} {arXiv:1605.04928 [cond.mat-quant-gas]}
  \BibitemShut {NoStop}%
\bibitem [{\citenamefont {Zou}\ \emph {et~al.}(2021)\citenamefont {Zou},
  \citenamefont {{Le Cerf}}, \citenamefont {{Bakkali-Hassani}}, \citenamefont
  {Maury}, \citenamefont {Chauveau}, \citenamefont {Castilho}, \citenamefont
  {{Saint-Jalm}}, \citenamefont {Nascimbene}, \citenamefont {Dalibard},\ and\
  \citenamefont {Beugnon}}]{Zou:2021opt}%
  \BibitemOpen
  \bibfield  {author} {\bibinfo {author} {\bibfnamefont {Y.-Q.}\ \bibnamefont
  {Zou}}, \bibinfo {author} {\bibfnamefont {{\'E}.}~\bibnamefont {{Le Cerf}}},
  \bibinfo {author} {\bibfnamefont {B.}~\bibnamefont {{Bakkali-Hassani}}},
  \bibinfo {author} {\bibfnamefont {C.}~\bibnamefont {Maury}}, \bibinfo
  {author} {\bibfnamefont {G.}~\bibnamefont {Chauveau}}, \bibinfo {author}
  {\bibfnamefont {P.~C.~M.}\ \bibnamefont {Castilho}}, \bibinfo {author}
  {\bibfnamefont {R.}~\bibnamefont {{Saint-Jalm}}}, \bibinfo {author}
  {\bibfnamefont {S.}~\bibnamefont {Nascimbene}}, \bibinfo {author}
  {\bibfnamefont {J.}~\bibnamefont {Dalibard}},\ and\ \bibinfo {author}
  {\bibfnamefont {J.}~\bibnamefont {Beugnon}},\ }\bibfield  {title} {\bibinfo
  {title} {{Optical control of the density and spin spatial profiles of a
  planar Bose gas}},\ }\href {https://doi.org/10.1088/1361-6455/abf298}
  {\bibfield  {journal} {\bibinfo  {journal} {J. Phys. B: At. Mol. Opt. Phys.}\
  }\textbf {\bibinfo {volume} {54}},\ \bibinfo {pages} {08LT01} (\bibinfo
  {year} {2021})},\ \Eprint {https://arxiv.org/abs/2102.05492}
  {arXiv:2102.05492 [cond-mat.quant-gas]} \BibitemShut {NoStop}%
\bibitem [{\citenamefont {Braden}\ \emph
  {et~al.}(2019{\natexlab{b}})\citenamefont {Braden}, \citenamefont {Johnson},
  \citenamefont {Peiris}, \citenamefont {Pontzen},\ and\ \citenamefont
  {Weinfurtner}}]{Braden:2018tky}%
  \BibitemOpen
  \bibfield  {author} {\bibinfo {author} {\bibfnamefont {J.}~\bibnamefont
  {Braden}}, \bibinfo {author} {\bibfnamefont {M.~C.}\ \bibnamefont {Johnson}},
  \bibinfo {author} {\bibfnamefont {H.~V.}\ \bibnamefont {Peiris}}, \bibinfo
  {author} {\bibfnamefont {A.}~\bibnamefont {Pontzen}},\ and\ \bibinfo {author}
  {\bibfnamefont {S.}~\bibnamefont {Weinfurtner}},\ }\bibfield  {title}
  {\bibinfo {title} {{New Semiclassical Picture of Vacuum Decay}},\ }\href
  {https://doi.org/10.1103/PhysRevLett.123.031601} {\bibfield  {journal}
  {\bibinfo  {journal} {Phys. Rev. Lett.}\ }\textbf {\bibinfo {volume} {123}},\
  \bibinfo {pages} {031601} (\bibinfo {year} {2019}{\natexlab{b}})},\ \bibinfo
  {note} {[Erratum: Phys.Rev.Lett. 129, 059901(E) (2022)]},\ \Eprint
  {https://arxiv.org/abs/1806.06069} {arXiv:1806.06069 [hep-th]} \BibitemShut
  {NoStop}%
\bibitem [{\citenamefont {Khlebnikov}\ and\ \citenamefont
  {Tkachev}(1996)}]{Khlebnikov:1996mc}%
  \BibitemOpen
  \bibfield  {author} {\bibinfo {author} {\bibfnamefont {S.~Y.}\ \bibnamefont
  {Khlebnikov}}\ and\ \bibinfo {author} {\bibfnamefont {I.~I.}\ \bibnamefont
  {Tkachev}},\ }\bibfield  {title} {\bibinfo {title} {{Classical decay of
  inflaton}},\ }\href {https://doi.org/10.1103/PhysRevLett.77.219} {\bibfield
  {journal} {\bibinfo  {journal} {Phys. Rev. Lett.}\ }\textbf {\bibinfo
  {volume} {77}},\ \bibinfo {pages} {219} (\bibinfo {year} {1996})},\ \Eprint
  {https://arxiv.org/abs/hep-ph/9603378} {arXiv:hep-ph/9603378} \BibitemShut
  {NoStop}%
\bibitem [{\citenamefont {Rajantie}\ \emph {et~al.}(2001)\citenamefont
  {Rajantie}, \citenamefont {Saffin},\ and\ \citenamefont
  {Copeland}}]{Rajantie:2000nj}%
  \BibitemOpen
  \bibfield  {author} {\bibinfo {author} {\bibfnamefont {A.}~\bibnamefont
  {Rajantie}}, \bibinfo {author} {\bibfnamefont {P.~M.}\ \bibnamefont
  {Saffin}},\ and\ \bibinfo {author} {\bibfnamefont {E.~J.}\ \bibnamefont
  {Copeland}},\ }\bibfield  {title} {\bibinfo {title} {{Electroweak preheating
  on a lattice}},\ }\href {https://doi.org/10.1103/PhysRevD.63.123512}
  {\bibfield  {journal} {\bibinfo  {journal} {Phys. Rev. D}\ }\textbf {\bibinfo
  {volume} {63}},\ \bibinfo {pages} {123512} (\bibinfo {year} {2001})},\
  \Eprint {https://arxiv.org/abs/hep-ph/0012097} {arXiv:hep-ph/0012097}
  \BibitemShut {NoStop}%
\bibitem [{\citenamefont {Garcia-Bellido}\ and\ \citenamefont
  {Figueroa}(2007)}]{Garcia-Bellido:2007nns}%
  \BibitemOpen
  \bibfield  {author} {\bibinfo {author} {\bibfnamefont {J.}~\bibnamefont
  {Garcia-Bellido}}\ and\ \bibinfo {author} {\bibfnamefont {D.~G.}\
  \bibnamefont {Figueroa}},\ }\bibfield  {title} {\bibinfo {title} {{A
  stochastic background of gravitational waves from hybrid preheating}},\
  }\href {https://doi.org/10.1103/PhysRevLett.98.061302} {\bibfield  {journal}
  {\bibinfo  {journal} {Phys. Rev. Lett.}\ }\textbf {\bibinfo {volume} {98}},\
  \bibinfo {pages} {061302} (\bibinfo {year} {2007})},\ \Eprint
  {https://arxiv.org/abs/astro-ph/0701014} {arXiv:astro-ph/0701014}
  \BibitemShut {NoStop}%
\bibitem [{\citenamefont {Amin}\ \emph {et~al.}(2012)\citenamefont {Amin},
  \citenamefont {Easther}, \citenamefont {Finkel}, \citenamefont {Flauger},\
  and\ \citenamefont {Hertzberg}}]{Amin:2011hj}%
  \BibitemOpen
  \bibfield  {author} {\bibinfo {author} {\bibfnamefont {M.~A.}\ \bibnamefont
  {Amin}}, \bibinfo {author} {\bibfnamefont {R.}~\bibnamefont {Easther}},
  \bibinfo {author} {\bibfnamefont {H.}~\bibnamefont {Finkel}}, \bibinfo
  {author} {\bibfnamefont {R.}~\bibnamefont {Flauger}},\ and\ \bibinfo {author}
  {\bibfnamefont {M.~P.}\ \bibnamefont {Hertzberg}},\ }\bibfield  {title}
  {\bibinfo {title} {{Oscillons After Inflation}},\ }\href
  {https://doi.org/10.1103/PhysRevLett.108.241302} {\bibfield  {journal}
  {\bibinfo  {journal} {Phys. Rev. Lett.}\ }\textbf {\bibinfo {volume} {108}},\
  \bibinfo {pages} {241302} (\bibinfo {year} {2012})},\ \Eprint
  {https://arxiv.org/abs/1106.3335} {arXiv:1106.3335 [astro-ph.CO]}
  \BibitemShut {NoStop}%
\bibitem [{\citenamefont {Clough}\ \emph {et~al.}(2017)\citenamefont {Clough},
  \citenamefont {Lim}, \citenamefont {DiNunno}, \citenamefont {Fischler},
  \citenamefont {Flauger},\ and\ \citenamefont {Paban}}]{Clough:2016ymm}%
  \BibitemOpen
  \bibfield  {author} {\bibinfo {author} {\bibfnamefont {K.}~\bibnamefont
  {Clough}}, \bibinfo {author} {\bibfnamefont {E.~A.}\ \bibnamefont {Lim}},
  \bibinfo {author} {\bibfnamefont {B.~S.}\ \bibnamefont {DiNunno}}, \bibinfo
  {author} {\bibfnamefont {W.}~\bibnamefont {Fischler}}, \bibinfo {author}
  {\bibfnamefont {R.}~\bibnamefont {Flauger}},\ and\ \bibinfo {author}
  {\bibfnamefont {S.}~\bibnamefont {Paban}},\ }\bibfield  {title} {\bibinfo
  {title} {{Robustness of Inflation to Inhomogeneous Initial Conditions}},\
  }\href {https://doi.org/10.1088/1475-7516/2017/09/025} {\bibfield  {journal}
  {\bibinfo  {journal} {JCAP}\ }\textbf {\bibinfo {volume} {09}}\bibfield
  {number} {\bibinfo  {number} { (2017)},\ \bibinfo {pages} {025}},\ }\Eprint
  {https://arxiv.org/abs/1608.04408} {arXiv:1608.04408 [hep-th]} \BibitemShut
  {NoStop}%
\bibitem [{\citenamefont {Blakie}\ \emph {et~al.}(2008)\citenamefont {Blakie},
  \citenamefont {Bradley}, \citenamefont {Davis}, \citenamefont {Ballagh},\
  and\ \citenamefont {Gardiner}}]{Blakie:2008vka}%
  \BibitemOpen
  \bibfield  {author} {\bibinfo {author} {\bibfnamefont {P.~B.}\ \bibnamefont
  {Blakie}}, \bibinfo {author} {\bibfnamefont {A.~S.}\ \bibnamefont {Bradley}},
  \bibinfo {author} {\bibfnamefont {M.~J.}\ \bibnamefont {Davis}}, \bibinfo
  {author} {\bibfnamefont {R.~J.}\ \bibnamefont {Ballagh}},\ and\ \bibinfo
  {author} {\bibfnamefont {C.~W.}\ \bibnamefont {Gardiner}},\ }\bibfield
  {title} {\bibinfo {title} {{Dynamics and statistical mechanics of ultra-cold
  Bose gases using c-field techniques}},\ }\href
  {https://doi.org/10.1080/00018730802564254} {\bibfield  {journal} {\bibinfo
  {journal} {Adv. Phys.}\ }\textbf {\bibinfo {volume} {57}},\ \bibinfo {pages}
  {363} (\bibinfo {year} {2008})},\ \Eprint {https://arxiv.org/abs/0809.1487}
  {arXiv:0809.1487 [cond-mat.stat-mech]} \BibitemShut {NoStop}%
\bibitem [{\citenamefont {Braden}\ \emph {et~al.}(2023)\citenamefont {Braden},
  \citenamefont {Johnson}, \citenamefont {Peiris}, \citenamefont {Pontzen},\
  and\ \citenamefont {Weinfurtner}}]{Braden:2022odm}%
  \BibitemOpen
  \bibfield  {author} {\bibinfo {author} {\bibfnamefont {J.}~\bibnamefont
  {Braden}}, \bibinfo {author} {\bibfnamefont {M.~C.}\ \bibnamefont {Johnson}},
  \bibinfo {author} {\bibfnamefont {H.~V.}\ \bibnamefont {Peiris}}, \bibinfo
  {author} {\bibfnamefont {A.}~\bibnamefont {Pontzen}},\ and\ \bibinfo {author}
  {\bibfnamefont {S.}~\bibnamefont {Weinfurtner}},\ }\bibfield  {title}
  {\bibinfo {title} {{Mass renormalization in lattice simulations of false
  vacuum decay}},\ }\href {https://doi.org/10.1103/PhysRevD.107.083509}
  {\bibfield  {journal} {\bibinfo  {journal} {Phys. Rev. D}\ }\textbf {\bibinfo
  {volume} {107}},\ \bibinfo {pages} {083509} (\bibinfo {year} {2023})},\
  \Eprint {https://arxiv.org/abs/2204.11867} {arXiv:2204.11867 [hep-th]}
  \BibitemShut {NoStop}%
\bibitem [{\citenamefont {Lysebo}\ and\ \citenamefont
  {Veseth}(2010)}]{Lysebo:2010fesh}%
  \BibitemOpen
  \bibfield  {author} {\bibinfo {author} {\bibfnamefont {M.}~\bibnamefont
  {Lysebo}}\ and\ \bibinfo {author} {\bibfnamefont {L.}~\bibnamefont
  {Veseth}},\ }\bibfield  {title} {\bibinfo {title} {{Feshbach resonances and
  transition rates for cold homonuclear collisions between ${}^{39}\mathrm{K}$
  and ${}^{41}\mathrm{K}$ atoms}},\ }\href
  {https://doi.org/10.1103/PhysRevA.81.032702} {\bibfield  {journal} {\bibinfo
  {journal} {Phys. Rev. A}\ }\textbf {\bibinfo {volume} {81}},\ \bibinfo
  {pages} {032702} (\bibinfo {year} {2010})}\BibitemShut {NoStop}%
\bibitem [{\citenamefont {Karailiev}\ \emph {et~al.}(2024)\citenamefont
  {Karailiev}, \citenamefont {Gazo}, \citenamefont {Ga{\l{}}ka}, \citenamefont
  {Eigen}, \citenamefont {Satoor},\ and\ \citenamefont
  {Hadzibabic}}]{Karailiev:2024sxs}%
  \BibitemOpen
  \bibfield  {author} {\bibinfo {author} {\bibfnamefont {A.}~\bibnamefont
  {Karailiev}}, \bibinfo {author} {\bibfnamefont {M.}~\bibnamefont {Gazo}},
  \bibinfo {author} {\bibfnamefont {M.}~\bibnamefont {Ga{\l{}}ka}}, \bibinfo
  {author} {\bibfnamefont {C.}~\bibnamefont {Eigen}}, \bibinfo {author}
  {\bibfnamefont {T.}~\bibnamefont {Satoor}},\ and\ \bibinfo {author}
  {\bibfnamefont {Z.}~\bibnamefont {Hadzibabic}},\ }\bibfield  {title}
  {\bibinfo {title} {{Observation of an Inverse Turbulent-Wave Cascade in a
  Driven Quantum Gas}},\ }\href
  {https://doi.org/10.1103/PhysRevLett.133.243402} {\bibfield  {journal}
  {\bibinfo  {journal} {Phys. Rev. Lett.}\ }\textbf {\bibinfo {volume} {133}},\
  \bibinfo {pages} {243402} (\bibinfo {year} {2024})},\ \Eprint
  {https://arxiv.org/abs/2405.01537} {arXiv:2405.01537 [cond-mat.quant-gas]}
  \BibitemShut {NoStop}%
\bibitem [{\citenamefont {Choi}\ \emph {et~al.}(1998)\citenamefont {Choi},
  \citenamefont {Morgan},\ and\ \citenamefont {Burnett}}]{Choi:1998ia}%
  \BibitemOpen
  \bibfield  {author} {\bibinfo {author} {\bibfnamefont {S.}~\bibnamefont
  {Choi}}, \bibinfo {author} {\bibfnamefont {S.~A.}\ \bibnamefont {Morgan}},\
  and\ \bibinfo {author} {\bibfnamefont {K.}~\bibnamefont {Burnett}},\
  }\bibfield  {title} {\bibinfo {title} {{Phenomenological damping in trapped
  atomic Bose-Einstein condensates}},\ }\href
  {https://doi.org/10.1103/PhysRevA.57.4057} {\bibfield  {journal} {\bibinfo
  {journal} {Phys. Rev. A}\ }\textbf {\bibinfo {volume} {57}},\ \bibinfo
  {pages} {4057} (\bibinfo {year} {1998})},\ \Eprint
  {https://arxiv.org/abs/quant-ph/9801064} {arXiv:quant-ph/9801064}
  \BibitemShut {NoStop}%
\bibitem [{\citenamefont {Pethick}\ and\ \citenamefont
  {Smith}(2008)}]{Pethick:2008bec}%
  \BibitemOpen
  \bibfield  {author} {\bibinfo {author} {\bibfnamefont {C.~J.}\ \bibnamefont
  {Pethick}}\ and\ \bibinfo {author} {\bibfnamefont {H.}~\bibnamefont
  {Smith}},\ }\href {https://doi.org/10.1017/CBO9780511802850} {\emph {\bibinfo
  {title} {{Bose-Einstein Condensation in Dilute Gases}}}},\ \bibinfo {edition}
  {2nd}\ ed.\ (\bibinfo  {publisher} {Cambridge University Press},\ \bibinfo
  {year} {2008})\BibitemShut {NoStop}%
\bibitem [{\citenamefont {Harris}\ \emph {et~al.}(2020)\citenamefont {Harris},
  \citenamefont {Millman}, \citenamefont {{van der Walt}}, \citenamefont
  {Gommers}, \citenamefont {Virtanen}, \citenamefont {Cournapeau},
  \citenamefont {Wieser}, \citenamefont {Taylor}, \citenamefont {Berg},
  \citenamefont {Smith}, \citenamefont {Kern}, \citenamefont {Picus},
  \citenamefont {Hoyer}, \citenamefont {{van Kerkwijk}}, \citenamefont {Brett},
  \citenamefont {Haldane}, \citenamefont {del R{\'{i}}o}, \citenamefont
  {Wiebe}, \citenamefont {Peterson}, \citenamefont {G{\'{e}}rard-Marchant},
  \citenamefont {Sheppard}, \citenamefont {Reddy}, \citenamefont {Weckesser},
  \citenamefont {Abbasi}, \citenamefont {Gohlke},\ and\ \citenamefont
  {Oliphant}}]{Harris:2020xlr}%
  \BibitemOpen
  \bibfield  {author} {\bibinfo {author} {\bibfnamefont {C.~R.}\ \bibnamefont
  {Harris}}, \bibinfo {author} {\bibfnamefont {K.~J.}\ \bibnamefont {Millman}},
  \bibinfo {author} {\bibfnamefont {S.~J.}\ \bibnamefont {{van der Walt}}},
  \bibinfo {author} {\bibfnamefont {R.}~\bibnamefont {Gommers}}, \bibinfo
  {author} {\bibfnamefont {P.}~\bibnamefont {Virtanen}}, \bibinfo {author}
  {\bibfnamefont {D.}~\bibnamefont {Cournapeau}}, \bibinfo {author}
  {\bibfnamefont {E.}~\bibnamefont {Wieser}}, \bibinfo {author} {\bibfnamefont
  {J.}~\bibnamefont {Taylor}}, \bibinfo {author} {\bibfnamefont
  {S.}~\bibnamefont {Berg}}, \bibinfo {author} {\bibfnamefont {N.~J.}\
  \bibnamefont {Smith}}, \bibinfo {author} {\bibfnamefont {R.}~\bibnamefont
  {Kern}}, \bibinfo {author} {\bibfnamefont {M.}~\bibnamefont {Picus}},
  \bibinfo {author} {\bibfnamefont {S.}~\bibnamefont {Hoyer}}, \bibinfo
  {author} {\bibfnamefont {M.~H.}\ \bibnamefont {{van Kerkwijk}}}, \bibinfo
  {author} {\bibfnamefont {M.}~\bibnamefont {Brett}}, \bibinfo {author}
  {\bibfnamefont {A.}~\bibnamefont {Haldane}}, \bibinfo {author} {\bibfnamefont
  {J.~F.}\ \bibnamefont {del R{\'{i}}o}}, \bibinfo {author} {\bibfnamefont
  {M.}~\bibnamefont {Wiebe}}, \bibinfo {author} {\bibfnamefont
  {P.}~\bibnamefont {Peterson}}, \bibinfo {author} {\bibfnamefont
  {P.}~\bibnamefont {G{\'{e}}rard-Marchant}}, \bibinfo {author} {\bibfnamefont
  {K.}~\bibnamefont {Sheppard}}, \bibinfo {author} {\bibfnamefont
  {T.}~\bibnamefont {Reddy}}, \bibinfo {author} {\bibfnamefont
  {W.}~\bibnamefont {Weckesser}}, \bibinfo {author} {\bibfnamefont
  {H.}~\bibnamefont {Abbasi}}, \bibinfo {author} {\bibfnamefont
  {C.}~\bibnamefont {Gohlke}},\ and\ \bibinfo {author} {\bibfnamefont {T.~E.}\
  \bibnamefont {Oliphant}},\ }\bibfield  {title} {\bibinfo {title} {{Array
  programming with NumPy}},\ }\href {https://doi.org/10.1038/s41586-020-2649-2}
  {\bibfield  {journal} {\bibinfo  {journal} {Nature}\ }\textbf {\bibinfo
  {volume} {585}},\ \bibinfo {pages} {357} (\bibinfo {year} {2020})},\ \Eprint
  {https://arxiv.org/abs/2006.10256} {arXiv:2006.10256 [cs.MS]} \BibitemShut
  {NoStop}%
\bibitem [{\citenamefont {Virtanen}\ \emph {et~al.}(2020)\citenamefont
  {Virtanen}, \citenamefont {Gommers}, \citenamefont {Oliphant}, \citenamefont
  {Haberland}, \citenamefont {Reddy}, \citenamefont {Cournapeau}, \citenamefont
  {Burovski}, \citenamefont {Peterson}, \citenamefont {Weckesser},
  \citenamefont {Bright}, \citenamefont {{van der Walt}}, \citenamefont
  {Brett}, \citenamefont {Wilson}, \citenamefont {Millman}, \citenamefont
  {Mayorov}, \citenamefont {Nelson}, \citenamefont {Jones}, \citenamefont
  {Kern}, \citenamefont {Larson}, \citenamefont {Carey}, \citenamefont {Polat},
  \citenamefont {Feng}, \citenamefont {Moore}, \citenamefont {VanderPlas},
  \citenamefont {Laxalde}, \citenamefont {Perktold}, \citenamefont {Cimrman},
  \citenamefont {Henriksen}, \citenamefont {Quintero}, \citenamefont {Harris},
  \citenamefont {Archibald}, \citenamefont {Ribeiro}, \citenamefont
  {Pedregosa}, \citenamefont {{van Mulbregt}},\ and\ \citenamefont {{SciPy 1.0
  Contributors}}}]{Virtanen:2019joe}%
  \BibitemOpen
  \bibfield  {author} {\bibinfo {author} {\bibfnamefont {P.}~\bibnamefont
  {Virtanen}}, \bibinfo {author} {\bibfnamefont {R.}~\bibnamefont {Gommers}},
  \bibinfo {author} {\bibfnamefont {T.~E.}\ \bibnamefont {Oliphant}}, \bibinfo
  {author} {\bibfnamefont {M.}~\bibnamefont {Haberland}}, \bibinfo {author}
  {\bibfnamefont {T.}~\bibnamefont {Reddy}}, \bibinfo {author} {\bibfnamefont
  {D.}~\bibnamefont {Cournapeau}}, \bibinfo {author} {\bibfnamefont
  {E.}~\bibnamefont {Burovski}}, \bibinfo {author} {\bibfnamefont
  {P.}~\bibnamefont {Peterson}}, \bibinfo {author} {\bibfnamefont
  {W.}~\bibnamefont {Weckesser}}, \bibinfo {author} {\bibfnamefont
  {J.}~\bibnamefont {Bright}}, \bibinfo {author} {\bibfnamefont {S.~J.}\
  \bibnamefont {{van der Walt}}}, \bibinfo {author} {\bibfnamefont
  {M.}~\bibnamefont {Brett}}, \bibinfo {author} {\bibfnamefont
  {J.}~\bibnamefont {Wilson}}, \bibinfo {author} {\bibfnamefont {K.~J.}\
  \bibnamefont {Millman}}, \bibinfo {author} {\bibfnamefont {N.}~\bibnamefont
  {Mayorov}}, \bibinfo {author} {\bibfnamefont {A.~R.~J.}\ \bibnamefont
  {Nelson}}, \bibinfo {author} {\bibfnamefont {E.}~\bibnamefont {Jones}},
  \bibinfo {author} {\bibfnamefont {R.}~\bibnamefont {Kern}}, \bibinfo {author}
  {\bibfnamefont {E.}~\bibnamefont {Larson}}, \bibinfo {author} {\bibfnamefont
  {C.~J.}\ \bibnamefont {Carey}}, \bibinfo {author} {\bibfnamefont
  {{\.I}.}~\bibnamefont {Polat}}, \bibinfo {author} {\bibfnamefont
  {Y.}~\bibnamefont {Feng}}, \bibinfo {author} {\bibfnamefont {E.~W.}\
  \bibnamefont {Moore}}, \bibinfo {author} {\bibfnamefont {J.}~\bibnamefont
  {VanderPlas}}, \bibinfo {author} {\bibfnamefont {D.}~\bibnamefont {Laxalde}},
  \bibinfo {author} {\bibfnamefont {J.}~\bibnamefont {Perktold}}, \bibinfo
  {author} {\bibfnamefont {R.}~\bibnamefont {Cimrman}}, \bibinfo {author}
  {\bibfnamefont {I.}~\bibnamefont {Henriksen}}, \bibinfo {author}
  {\bibfnamefont {E.~A.}\ \bibnamefont {Quintero}}, \bibinfo {author}
  {\bibfnamefont {C.~R.}\ \bibnamefont {Harris}}, \bibinfo {author}
  {\bibfnamefont {A.~M.}\ \bibnamefont {Archibald}}, \bibinfo {author}
  {\bibfnamefont {A.~H.}\ \bibnamefont {Ribeiro}}, \bibinfo {author}
  {\bibfnamefont {F.}~\bibnamefont {Pedregosa}}, \bibinfo {author}
  {\bibfnamefont {P.}~\bibnamefont {{van Mulbregt}}},\ and\ \bibinfo {author}
  {\bibnamefont {{SciPy 1.0 Contributors}}},\ }\bibfield  {title} {\bibinfo
  {title} {{SciPy 1.0--Fundamental Algorithms for Scientific Computing in
  Python}},\ }\href {https://doi.org/10.1038/s41592-019-0686-2} {\bibfield
  {journal} {\bibinfo  {journal} {Nature Meth.}\ }\textbf {\bibinfo {volume}
  {17}},\ \bibinfo {pages} {261} (\bibinfo {year} {2020})},\ \Eprint
  {https://arxiv.org/abs/1907.10121} {arXiv:1907.10121 [cs.MS]} \BibitemShut
  {NoStop}%
\bibitem [{\citenamefont {Hunter}(2007)}]{Hunter:2007ouj}%
  \BibitemOpen
  \bibfield  {author} {\bibinfo {author} {\bibfnamefont {J.~D.}\ \bibnamefont
  {Hunter}},\ }\bibfield  {title} {\bibinfo {title} {{Matplotlib: A 2D Graphics
  Environment}},\ }\href {https://doi.org/10.1109/MCSE.2007.55} {\bibfield
  {journal} {\bibinfo  {journal} {Comput. Sci. Eng.}\ }\textbf {\bibinfo
  {volume} {9}},\ \bibinfo {pages} {90} (\bibinfo {year} {2007})}\BibitemShut
  {NoStop}%
\bibitem [{\citenamefont {Jenkins}\ \emph {et~al.}(2025)\citenamefont
  {Jenkins}, \citenamefont {Peiris},\ and\ \citenamefont
  {Pontzen}}]{2504.02829-data}%
  \BibitemOpen
  \bibfield  {author} {\bibinfo {author} {\bibfnamefont {A.~C.}\ \bibnamefont
  {Jenkins}}, \bibinfo {author} {\bibfnamefont {H.~V.}\ \bibnamefont
  {Peiris}},\ and\ \bibinfo {author} {\bibfnamefont {A.}~\bibnamefont
  {Pontzen}},\ }\href {https://doi.org/10.5281/zenodo.15783498} {\bibinfo
  {title} {{Simulation data for arXiv:2504.02829}}} (\bibinfo {year} {2025}),\
  \bibinfo {note}
  {\url{https://github.com/alex-c-jenkins/2504.02829-data}}\BibitemShut
  {NoStop}%
\end{thebibliography}%
\end{document}